\documentclass[11pt]{article}
\usepackage[left=2.5cm,right=2.5cm,top=2.5cm,bottom=2.5cm,a4paper]{geometry}
\usepackage{graphicx}
\usepackage{amsmath,amssymb}
\usepackage{physics}
\usepackage[affil-it]{authblk}
\usepackage{bm}
\usepackage{cite}
\usepackage{subcaption}
\usepackage{url}
\usepackage{autobreak}
\usepackage{placeins}

\newcommand{\gev}{\,{\rm GeV}}

\newcommand{\oder}[1]{\mathcal{O}(#1)}

\newcommand{\SthreeDeformed}{3\times10^{2}}   
\newcommand{\SthreeRef}{6\times10^{2}}  

\title{Nonperturbative functional renormalization group for Higgs-singlet models with physics‑informed neural networks}
\author{Norimi Yokozaki$^{(a)}$\thanks{n.yokozaki@gmail.com}}
\affil{{\small (a) Zhejiang Institute of Modern Physics and Department of Physics, Zhejiang University, Hangzhou, Zhejiang 310027, China}}
\date{}

\begin{document}
	\maketitle

\begin{abstract}
We develop a nonperturbative functional renormalization group framework within the LPA' to solve the Wetterich flow equation for the $Z_2$-symmetric real singlet extension of the Standard Model at finite temperature, without a low-order polynomial truncation of the loop corrections to the scale-dependent effective potential, using a physics-informed neural network (PINN) representation. In contrast to conventional truncations based on low-order field expansions, our hybrid tree-level--plus--neural-network ansatz yields a continuous, mesh-free description of the effective potential over the full relevant field and scale range. As a proof of concept, we apply the framework to the finite-temperature effective potential relevant to the electroweak phase transition in this model: one-dimensional field-space slices at two benchmark temperatures, and a two-dimensional reconstruction at $T=100$ GeV from which a two-step bounce action is extracted. The renormalization group flow is implemented in a multi-domain setup in scale and field space with derivative matching conditions that ensure smoothness and numerical stability. Gauge and Yukawa sectors are incorporated via independently computed perturbative two-loop running couplings, and anomalous dimensions are included in the flow at the LPA' level. We benchmark the network against resummed perturbation theory and against a grid-based relaxation solver of the same equation. Finally, we introduce a soft consistency constraint that keeps the solution close, in sign and order of magnitude, to perturbation theory across the field-space domain. We find this constraint necessary, rather than merely helpful, for selecting a physically sensible solution of the flow equation. The converged result nevertheless retains a residual dependence on this guidance -- through hand-tuned weights and an analytic thermal target -- which we identify as the central open problem for mesh-free FRG treatments of this kind.
\end{abstract}

\section{Introduction}
\label{sec:intro}
	The Standard Model (SM) of particle physics, completed by the discovery of
	the Higgs boson~\cite{Aad:2012tfa,Chatrchyan:2012xdj}, describes all observed collider phenomena with
	remarkable accuracy and remains consistent with essentially every precision
	test performed to date.

	Despite this success, the SM cannot be the complete theory of nature. It
	does not account for neutrino masses, dark matter, or the origin of the
	observed cosmic matter--antimatter asymmetry, leaving several central
	problems of cosmology and particle physics unresolved.

	Among these, the baryon asymmetry of the universe (BAU) is a particularly
	sharp challenge: the SM cannot generate the observed asymmetry through the
	Sakharov conditions in sufficient
	quantity~\cite{Sakharov:1967dj,Kuzmin:1985mm,Morrissey:2012db,vandeVis:2025bkl}. This
	motivates extending the scalar sector of the SM in ways that can support a
	viable baryogenesis mechanism while remaining compatible with current
	collider and dark matter constraints.

	Electroweak baryogenesis (EWBG) offers an attractive mechanism to generate
	the BAU at the electroweak scale, but it requires two separate ingredients
	absent from the SM: a strongly first-order EWPT and new sources of CP
	violation. In the SM itself neither is realized -- lattice studies show
	that the electroweak transition is a smooth crossover rather than a
	first-order transition~\cite{Kajantie:1996mn}, and the CP violation of the
	CKM matrix is far too small to seed the observed asymmetry -- so EWBG is
	only a viable option once the scalar sector is extended. Extending the SM
	with an additional scalar, most simply a real singlet, is one of the
	best-studied ways to restore a strongly first-order EWPT: the extra
	scalar can modify the shape of the finite-temperature potential enough to
	generate a barrier between the symmetric and broken phases, and such
	singlet extensions have been extensively analyzed in connection with dark
	matter, vacuum stability, and the EWPT, both for the real, $Z_2$-symmetric
	case that we adopt below~\cite{Espinosa:1993bs,Profumo:2007wc,Ghorbani:2020xnv,Niemi:2021qvp,Niemi:2024xsm}
	and, once the singlet is complexified or coupled with additional CP-odd
	interactions, as a source of the CP violation needed to complete
	EWBG~\cite{Cline:2012hg,Grzadkowski:2018nbc}. Reliable computations of the
	finite-temperature effective potential are essential for establishing
	whether a given extension actually achieves a strongly first-order
	transition, but conventional perturbative techniques suffer from severe
	IR sensitivity and resummation issues near the critical temperature of a
	first-order EWPT.

	The functional renormalization group (FRG), formulated via the Wetterich
	equation, provides a non-perturbative tool which smoothly interpolates
	between the bare UV action and the full IR effective action, and is
	well suited to resolving these IR issues. However, in scalar extensions of
	the SM the effective potential depends on multiple field invariants,
	leading to a nonlinear PDE in more than one variable, which is technically
	demanding to solve by standard finite-difference or pseudo-spectral
	methods. Most functional renormalization group studies of scalar
	extensions of the Standard Model rely on low-order polynomial truncations
	of the effective potential or grid-based discretizations in field space.

	Here we consider the real singlet scalar extension of the SM with $Z_2$
	symmetry as a concrete setting for this FRG treatment. Being $Z_2$-symmetric
	and real, the singlet introduces no new source of CP violation, so the
	model by itself cannot complete EWBG; we take the strongly first-order
	transition as the first step to be established quantitatively, using this
	model as a controlled testing ground for the finite-temperature
	effective-potential methodology, and leave the additional CP-violating
	sector required to generate the BAU to future extensions of the model.

	Physics-informed neural networks (PINNs)~\cite{Raissi:2019} parametrize the unknown function
	as a neural network and minimize the PDE residual via automatic
	differentiation. They are mesh-free and naturally handle multi-dimensional
	domains, making them a natural candidate for the multi-field FRG flow
	equations that are otherwise costly to solve. Applying this idea to the
	FRG flow equation itself is a recent development: a genuinely mesh-free
	treatment, in which a neural network represents the field dependence
	continuously and the flow equation is enforced through automatic
	differentiation of the network output, has been demonstrated for a
	single-field $O(N)$-symmetric scalar theory~\cite{Yokota:2023pinn}, and a
	related neural-network parametrization of the field- and
	scale-dependent potential for the $O(N)$ model at finite temperature has
	also been proposed, though there the network is trained against a
	finite-difference-discretized residual evaluated on a fixed
	field-and-RG-time grid rather than a continuous PDE
	residual~\cite{Tan:2026frg}. Extending a genuinely continuous,
	autodiff-based PINN treatment to a multi-field, finite-temperature
	effective potential of a phenomenologically motivated BSM extension --
	as required for a first-order EWPT analysis -- remains unexplored. In
	this paper we apply PINNs to the FRG equation of the $Z_2$-symmetric
	singlet extension of the SM with finite-temperature contributions and
	SM gauge/Yukawa loops. Our aim here is methodological: to establish the
	end-to-end pipeline -- from UV couplings, through the mesh-free flow, to a
	finite-temperature effective potential and a bounce action -- as a proof
	of concept, rather than to deliver a full temperature scan or a converged
	determination of the transition strength at this benchmark.

	The remainder of this paper is organized as follows. Sec.~\ref{sec:wetterich} sets up the
	Wetterich equation for the $Z_2$ singlet extension and its dimensionless
	LPA$'$ truncation. Sec.~\ref{sec:ewpt} reviews the resulting two-step electroweak
	phase transition and the critical and nucleation temperatures used to
	assess a strongly first-order transition. Sec.~\ref{sec:relax-benchmark} fixes the benchmark
	point and input parameters used throughout, including the one-loop
	Coleman--Weinberg matching at $k_{\rm IR}$ and the two-loop RGE running
	to $k_{\rm UV}$, and introduces a grid-based Newton--Krylov relaxation
	solver as a benchmark for the flow equation. It first checks the solver on
	a singlet-free reduction of the model -- the simplest, single-field version
	of the problem -- and then on the full two-field system, so as to test
	whether the observed convergence failure originates in the
	Higgs--singlet mass-eigenvalue mixing or is already present in the
	single-field loop-level flow equation. Sec.~\ref{sec:network}
	describes the PINN architecture, decomposition of the potential, loss
	function, and multi-block training scheme used to solve the same flow
	equation. Sec.~\ref{sec:results} presents the resulting PINN solutions, again starting from
	the singlet-free reduction before the full system, compares them
	with the perturbative and relaxation-based benchmarks, and discusses the
	practical role and limitations of the soft perturbative-consistency
	penalty used in the PINN loss. Sec.~\ref{sec:conclusions} summarizes our findings and
	outlines directions for future work, including the CP-violating
	extensions needed to complete EWBG.

\FloatBarrier
\section{Wetterich equation for the $Z_2$ singlet extension}
\label{sec:wetterich}

\FloatBarrier
\subsection{Model and truncation}
\label{sec:model}

We consider the scalar sector of the Standard Model extended by a real singlet
	scalar $S$ with a $Z_2$ symmetry,
	\begin{align}
		S \;\to\; -S .
	\end{align}
	In this subsection, we keep only scalar fluctuations and neglect gauge and
	Yukawa contributions.

	At the UV scale $k=\Lambda$, the $Z_2$-symmetric scalar potential is
	\begin{align}
		V_\Lambda(H,S)
		&=
		\mu_H^2 |H|^2 + \lambda_H |H|^4
		+ \frac{1}{2}\mu_S^2 S^2 + \frac{1}{4}\lambda_S S^4
		+ \frac{1}{2}\lambda_{HS}|H|^2 S^2 .
	\end{align}
	Introducing the field invariants
	\begin{align}
		\tilde\rho \equiv H^\dagger H,
		\qquad
		\tilde\sigma \equiv \frac{1}{2} S^2 ,
	\end{align}
	the potential can be written as
	\begin{align}
		V_\Lambda(\tilde\rho,\tilde\sigma)
		&=
		\mu_H^2 \tilde\rho + \lambda_H \tilde\rho^2
		+ \mu_S^2 \tilde\sigma + \lambda_S \tilde\sigma^2
		+ \lambda_{HS}\tilde\rho\tilde\sigma .
	\end{align}

	Within the derivative expansion at LPA$'$~\cite{Dupuis:2020fhq}, we employ the ansatz
	\begin{align}
		\Gamma_k[H,S]
		=
		\int_x
		\biggl[
		Z_{H,k}\,(\partial_\mu H)^\dagger (\partial_\mu H)
		+ \frac{1}{2} Z_{S,k}\,(\partial_\mu S)^2
		+ V_k(\tilde\rho,\tilde\sigma)
		\biggr] ,
	\end{align}
	where
	\begin{align}
		\int_x \equiv \int_0^{1/T} d\tau \int d^3x ,
		\qquad
		t \equiv \ln\frac{k}{\Lambda} ,
	\end{align}
	and the anomalous dimensions are defined by
	\begin{align}
		\eta_H \equiv - \partial_t \ln Z_{H,k},
		\qquad
		\eta_S \equiv - \partial_t \ln Z_{S,k} .
	\end{align}

	\subsection{Wetterich equation}
	\label{sec:wetterich-eq}

	The flow of the effective average action is governed by the Wetterich equation~\cite{Wetterich:1992yh,Berges:2000ew},
	\begin{align}
		\partial_t \Gamma_k
		=
		\frac{1}{2}\,\mathrm{STr}
		\left[
		\left(
		\Gamma_k^{(2)} + R_k
		\right)^{-1}
		\partial_t R_k
		\right] .
	\end{align}

	At finite temperature, the Euclidean zeroth component is replaced by the bosonic
	Matsubara frequencies,
	\begin{align}
		p_0 \to \omega_n = 2\pi n T,
		\qquad
		\int \frac{dp_0}{2\pi}
		\to
		T \sum_{n\in\mathbb{Z}} .
	\end{align}

	Using the Litim regulator~\cite{Litim:2001up},
		\begin{align}
		R_{H,k}(p)
		&=
		Z_{H,k}\,
		(k^2 - \bm{p}^2)\,
		\theta\!\left(k^2 - \bm{p}^2 \right),
		\\
		R_{S,k}(p)
		&=
		Z_{S,k}\,
		(k^2 - \bm{p}^2)\,
		\theta\!\left(k^2 - \bm{p}^2\right),
	\end{align}
	and evaluating the flow for homogeneous background
	fields, the momentum-shell integral is elementary and the Matsubara sum
	closes in terms of a $\coth$~\cite{Litim:2006ag}, giving the flow of the
	effective potential,
	\begin{align}
		\partial_t V_k(\tilde\rho,\tilde\sigma)
		&=
		\frac{k^4}{12\pi^2}
		\left[
		\frac{3}{\sqrt{1+\bar m_G^2(\tilde\rho,\tilde\sigma)}}
		\coth\!\left(
		\frac{\sqrt{1+\bar m_G^2(\tilde\rho,\tilde\sigma)}}{2\tau}
		\right)
		\right.
		\nonumber\\
		&\hspace{2cm}
		+
		\frac{1}{\sqrt{1+\bar m_+^2(\tilde\rho,\tilde\sigma)}}
		\coth\!\left(
		\frac{\sqrt{1+\bar m_+^2(\tilde\rho,\tilde\sigma)}}{2\tau}
		\right)
		\nonumber\\
		&\hspace{2cm}
		\left.
		+
		\frac{1}{\sqrt{1+\bar m_-^2(\tilde\rho,\tilde\sigma)}}
		\coth\!\left(
		\frac{\sqrt{1+\bar m_-^2(\tilde\rho,\tilde\sigma)}}{2\tau}
		\right)
		\right]
		\nonumber\\
		&\hspace{2cm}
		+ k^4\,E_{\rm ext},
	\end{align}
	where $\tau \equiv T/k$. Each bosonic mode $i$ in the bracket arises as
	\begin{align}
		\frac{1}{2}\int\!\frac{d^3p}{(2\pi)^3}\,T\!\sum_{n\in\mathbb{Z}}
		\frac{\partial_t R_{k}}{\omega_n^2+\bm p^2+R_k+\bar m_i^2 k^2}
		=
		\frac{k^4}{12\pi^2}\,\frac{1}{\sqrt{1+\bar m_i^2}}\,
		\coth\!\left(\frac{k\sqrt{1+\bar m_i^2}}{2T}\right) ,
		\label{eq:mode-contribution}
	\end{align}
	since the $\theta$-function collapses the momentum-shell integral to
	$\int_{|\bm p|<k}\!d^3p/(2\pi)^3=k^3/6\pi^2$ with $\partial_t R_k\to 2k^2$ inside it,
	while the Matsubara sum $T\sum_n(\omega_n^2+E_i^2)^{-1}=(2E_i)^{-1}\coth(E_i/2T)$,
	with $E_i=k\sqrt{1+\bar m_i^2}$, is elementary; the bracket sums this over the three
	Goldstone modes and the two radial eigenvalues.
	The term $E_{\rm ext}$ is the dimensionless sum of contributions from top quark and electroweak gauge boson loops (W and Z), evaluated within the same regulator scheme; it is given explicitly in Eq.~\eqref{eq:E-ext} below and enters the dimensionless flow~\eqref{eq:flow-dimless} directly, without the $k^4$ prefactor.
	Here, the three Goldstone modes contribute through
	\begin{align}
		\bar m_G^2(\tilde\rho,\tilde\sigma)
		=
		\frac{1}{k^2 Z_{H,k}}
		\frac{\partial V_k}{\partial \tilde\rho} ,
	\end{align}
	while $\bar m_\pm^2$ are the eigenvalues of the $2\times 2$ scalar radial mass matrix $\bar M^2$
	in the Higgs--singlet sector. Their dimensionless forms are given in Sec.~\ref{sec:dimless}.

	\subsection{Dimensionless form}
	\label{sec:dimless}

	We introduce the dimensionless renormalized variables
	\begin{align}
		u(\rho,\sigma)
		\equiv
		\frac{V_k(\tilde\rho,\tilde\sigma)}{k^4},
		\qquad
		\rho \equiv \frac{Z_{H,k}\tilde\rho}{k^2},
		\qquad
		\sigma \equiv \frac{Z_{S,k}\tilde\sigma}{k^2},
	\end{align}
	where the field invariants are normalized as $\tilde\rho = H^\dagger H = h^2/2$ and $\tilde\sigma = s^2/2$, with $h,s$ the physical (canonically normalized) radial fields and $\langle h\rangle = v = 246.2\gev$ at the electroweak minimum. The flow equation and the neural-network representation work throughout with the dimensionless $(\rho,\sigma)$; the physical field-space quantities of Sec.~\ref{sec:ewpt} (vacuum locations, $v_c$, and the bounce trajectory $\phi=(h,s)$) are quoted in the dimensionful invariants $(\tilde\rho,\tilde\sigma)$ in $\gev^2$.
	The UV potential becomes
	\begin{align}
		u_\Lambda(\rho,\sigma)
		&=
		a_H \rho + \lambda_H \rho^2
		+ a_S \sigma + \lambda_S \sigma^2
		+ \lambda_{HS} \rho \sigma,
		\label{eq:u-uv}
	\end{align}
	where $a_H = \mu_H^2/\Lambda^2$ and $a_S = \mu_S^2/\Lambda^2$.
	Then the flow equation becomes
	\begin{align}
		\partial_t u(\rho,\sigma)
		&=
		- 4\,u(\rho,\sigma)
		+ (2+\eta_H)\,\rho\,u_\rho
		+ (2+\eta_S)\,\sigma\,u_\sigma
		\nonumber \\
		&\quad
		+\frac{1}{12\pi^2}
		\left[
		\frac{3}{\sqrt{1+\bar m_G^2(\rho,\sigma)}}
		\coth\!\left(
		\frac{\sqrt{1+\bar m_G^2(\rho,\sigma)}}{2\tau}
		\right)
		\right.
		\nonumber\\
		&\quad
		+
		\frac{1}{\sqrt{1+\bar m_+^2(\rho,\sigma)}}
		\coth\!\left(
		\frac{\sqrt{1+\bar m_+^2(\rho,\sigma)}}{2\tau}
		\right)
		\nonumber\\
		&\quad
		\left.
		+
		\frac{1}{\sqrt{1+\bar m_-^2(\rho,\sigma)}}
		\coth\!\left(
		\frac{\sqrt{1+\bar m_-^2(\rho,\sigma)}}{2\tau}
		\right)
		\right]
		\nonumber\\
		&\quad + E_{\rm ext}(t,\tau, \rho)
		\label{eq:flow-dimless}
	\end{align}
	where
	\begin{align}
		u_\rho \equiv \frac{\partial u}{\partial \rho},
		\qquad
		u_\sigma \equiv \frac{\partial u}{\partial \sigma},
		\qquad
		u_{\rho\rho} \equiv \frac{\partial^2 u}{\partial \rho^2},
		\qquad
		u_{\sigma\sigma} \equiv \frac{\partial^2 u}{\partial \sigma^2},
		\qquad
		u_{\rho\sigma} \equiv \frac{\partial^2 u}{\partial \rho\,\partial \sigma} .
	\end{align}
	In the zero-temperature limit, $\tau \to 0$, one has $\coth(x/2\tau)\to 1$, and
	the above equation reduces to the usual $T=0$ flow.

	Explicitly,
	\begin{align}
		E_{\rm ext}(t,\tau,\rho)
		=
		\frac{1}{12\pi^2}
		\left[
		\frac{6}{\sqrt{1+\bar m_W^2}}
		\coth\!\left(\frac{\sqrt{1+\bar m_W^2}}{2\tau}\right)
		+
		\frac{3}{\sqrt{1+\bar m_Z^2}}
		\coth\!\left(\frac{\sqrt{1+\bar m_Z^2}}{2\tau}\right)
		\right.
		\nonumber\\
		\left.
		-\,
		\frac{12}{\sqrt{1+\bar m_t^2}}
		\tanh\!\left(\frac{\sqrt{1+\bar m_t^2}}{2\tau}\right)
		\right],
		\label{eq:E-ext}
	\end{align}
	with dimensionless tree-level masses
	$\bar m_W^2 = g_2^2\rho/2$, $\bar m_Z^2 = (g_1^2+g_2^2)\rho/2$, and
	$\bar m_t^2 = y_t^2\rho$ (evaluated with the running $g_1,g_2,y_t$ at
	scale $k$), and combined multiplicities $6$ and $3$ counting transverse
	and longitudinal polarizations of $W^\pm$ and $Z$ at equal mass (no
	separate Debye improvement of the longitudinal modes is applied inside
	the flow equation itself). The top quark enters with an overall minus
	sign and $\tanh$ rather than $\coth$, as required by Fermi--Dirac
	statistics; as $\tau\to0$, $\tanh\to1$ and $E_{\rm ext}$ reduces to the
	usual zero-temperature top and gauge-boson threshold functions.

	The dimensionless Goldstone mass is
	\begin{align}
		\bar m_G^2(\rho,\sigma) = u_\rho(\rho,\sigma) .
		\label{eq:mass-goldstone}
	\end{align}
	The dimensionless radial mass matrix is given by
	\begin{align}
		\bar M^2(\rho,\sigma)
		=
		\begin{pmatrix}
			u_\rho + 2\rho\,u_{\rho\rho}
			&
			2\sqrt{\rho\sigma}\,u_{\rho\sigma}
			\\
			2\sqrt{\rho\sigma}\,u_{\rho\sigma}
			&
			u_\sigma + 2\sigma\,u_{\sigma\sigma}
		\end{pmatrix} ,
		\label{eq:mass-matrix}
	\end{align}
	and its eigenvalues are
	\begin{align}
		\bar m_\pm^2(\rho,\sigma)
		=
		\frac{1}{2}
		\left[
		\bar M_{11}^2 + \bar M_{22}^2
		\pm
		\sqrt{
			\left(\bar M_{11}^2 - \bar M_{22}^2\right)^2
			+ 4\left(\bar M_{12}^2\right)^2
		}
		\right] .
		\label{eq:mass-eigen}
	\end{align}
	In our Higgs--singlet setup, the Higgs field is directly coupled to the top quark and electroweak gauge bosons, while the singlet is not.
	Consequently, the corresponding wave-function renormalizations $Z_{H,k}$ and $Z_{S,k}$ are expected to exhibit a markedly different scale dependence.
	Perturbatively, one finds that the anomalous dimension $\eta_H$ is generated already at one loop, whereas $\eta_S$ starts only at two loops.
	In the renormalized variables $(\rho,\sigma)$ the wave-function factors cancel identically from Eq.~\eqref{eq:mass-matrix}: inside the last integrated momentum shell the radial propagator is $Z_{i,k}k^2+\partial^2 V_k/\partial\phi_i\partial\phi_j$, and rescaling by $Z_{i,k}k^2$ turns the curvature of $V_k$ into exactly the curvature of $u$. Equation~\eqref{eq:mass-matrix} is therefore already the canonically normalized radial mass matrix, with the Higgs--singlet mixing $u_{\rho\sigma}$ fully retained through the eigenvalues $\bar m_\pm^2$; the difference $Z_{H,k}\neq Z_{S,k}$ does not enter it.\footnote{It would enter only through the $\eta_i$-dependent factor $(1-\eta_i/(d+2))$ that a fully consistent scale derivative of the $Z_{i,k}$-dressed regulator attaches to each threshold function -- there weighting a mixing-angle-dependent combination of $\eta_H$ and $\eta_S$. Our implementation evaluates the threshold functions at $\eta=0$; with $\eta_H\lesssim0.02$ and $\eta_S\sim10^{-4}$ this omission is at the sub-percent level.} What the LPA$'$ ansatz does drop is the field dependence of $Z_{H,k},Z_{S,k}$ and any off-diagonal kinetic term $Z_{\rho\sigma}\,\partial_\mu\rho\,\partial_\mu\sigma$ mixing the two radial modes.\footnote{The cancellation above is exact for the diagonal $Z_{H,k},Z_{S,k}$ of the ansatz, which the rescaling $\rho\propto Z_{H,k}\tilde\rho$, $\sigma\propto Z_{S,k}\tilde\sigma$ undoes. A retained $Z_{\rho\sigma}$ has no counterpart in that rescaling and would add off-diagonal corrections to $\bar M^2$.} The latter is generated by the portal coupling alongside $u_{\rho\sigma}$, but a constant $H$--$S$ kinetic mixing is forbidden by gauge invariance, so it is field-dependent (vanishing at the origin by the $Z_2$ symmetry) and parametrically of order $\lambda_{HS}^2/(16\pi^2)$ times a loop factor -- the same size as $\eta_S$ itself.

If one works in strict LPA rather than LPA$'$, one simply sets
\begin{align}
	\eta_H = \eta_S = 0 .
\end{align}
For the numerical calculations, we directly evaluate the dimensionless
potential $u(\rho,\sigma)$ by means of physics-informed neural
networks (PINNs), whereas the anomalous dimensions $\eta_H$ and
$\eta_S$ are taken from perturbative one- and two-loop results
(the leading contribution to $\eta_S$ appears only at two-loop order).

At one loop, the Higgs anomalous dimension in the SM is given by
\begin{align}
\gamma_H \equiv - \frac{1}{2}\partial_t \ln Z_H
\simeq
\frac{1}{16\pi^2}
\left(
3 y_t^2
- \frac{9}{4} g_2^2
- \frac{3}{4} g_1^2
\right),  \  \eta_H \simeq 2 \gamma_H \, ,
\end{align}
in the $\overline{\text{MS}}$ scheme with Landau gauge.
For the singlet $S$, $\gamma_S$ vanishes at one loop and is first generated at two loops through the portal and singlet quartic couplings; PyR@TE3~\cite{Sartore:2020gyh} gives
\begin{align}
\gamma_S \simeq
\frac{1}{(16\pi^2)^2}
\left(
\lambda_{HS}^2
+ 3 \lambda_{S}^2
\right), \  \eta_S \simeq 2 \gamma_S  \, .
\end{align}
Over the RG-time range used here $\eta_S\lesssim10^{-4}$, so its precise value (and its residual scheme dependence) has no visible effect on the results; it is retained only for completeness. Both the one-loop $\gamma_H$ and the two-loop $\gamma_S$ were obtained with PyR@TE3 for the model defined above; the full set of two-loop $\beta$ functions is collected in Appendix~\ref{app:rge}.

\FloatBarrier
\section{First-order electroweak phase transition}
\label{sec:ewpt}

\FloatBarrier
\subsection{Vacuum structure and two-step transitions}
\label{sec:vacuum}

In the $Z_2$-symmetric singlet extension, the zero-temperature potential
can support, in addition to the electroweak vacuum
$(\tilde\rho,\tilde\sigma) = (v^2/2,\,0)$, a singlet-only vacuum
$(\tilde\rho,\tilde\sigma) = (0,\,v_S^2/2)$ where the $Z_2$ symmetry is spontaneously
broken. At high temperature both minima are typically lifted relative to
the symmetric point $(\tilde\rho,\tilde\sigma)=(0,0)$, and depending on the size of
$\lambda_{HS}$ the thermal history can proceed either directly from the
symmetric phase to the electroweak vacuum (one-step transition), or via an
intermediate stage in which the singlet direction is populated first and
the system subsequently tunnels into the electroweak vacuum (two-step
transition)~\cite{Espinosa:1993bs,Ghorbani:2020xnv}.

Formally, a two-step history is characterized by a temperature range in
which the singlet-broken minimum is the global minimum of the finite-$T$
potential,
\begin{align}
	V(0,v_S;T) < V(v,0;T) ,
\end{align}
followed, at lower temperature, by a swap of the global minimum,
\begin{align}
	V(0,v_S;T) > V(v,0;T) ,
\end{align}
so that the electroweak vacuum eventually becomes the true ground state.
The nature of this vacuum swap --- and in particular whether the second
step proceeds via a strongly first-order transition --- is highly
sensitive to the shape of the potential barrier separating the two minima,
which in turn is controlled by the same nonperturbative infrared dynamics
that the FRG flow is designed to resolve. This is in contrast to one-loop
perturbative treatments, where the barrier is generated only by thermal
cubic terms and is therefore parametrically small unless the portal
coupling $\lambda_{HS}$ is sizable~\cite{Profumo:2007wc}. Two-loop perturbative studies
have shown that such corrections can shift the transition strength and
critical temperature by $\oder{20\%\text{--}50\%}$ relative to the
one-loop result in typical strong-transition benchmarks~\cite{Niemi:2021qvp}, which
motivates a fully nonperturbative treatment such as the one pursued here. A
nonperturbative lattice study of this model, based on a perturbatively
dimensionally-reduced three-dimensional effective theory, has recently mapped
the strongly first-order region of its parameter space and provides the natural
benchmark against which a fully nonperturbative effective potential should
ultimately be tested~\cite{Niemi:2024xsm}.

\FloatBarrier
\subsection{Critical and nucleation temperatures}
\label{sec:tc-tn}

For a given temperature $T$, the flow is integrated from the UV boundary down
to the infrared scale $k_{\rm IR}=150\gev$ (Sec.~\ref{sec:benchmark-point}),
yielding the flowed potential $u_{k_{\rm IR}}(\rho,\sigma;T)$ used to determine
the vacuum structure. For the vacuum structure and the bounce below we work with
the corresponding dimensionful potential
\begin{align}
	V_{k_{\rm IR}}(\tilde\rho,\tilde\sigma;T)
	\equiv
	k_{\rm IR}^4\,
	u_{k_{\rm IR}}\!\bigl(
	Z_{H,k_{\rm IR}}\tilde\rho/k_{\rm IR}^2,\;
	Z_{S,k_{\rm IR}}\tilde\sigma/k_{\rm IR}^2;\,T
	\bigr) ,
\end{align}
a function of the dimensionful invariants $(\tilde\rho,\tilde\sigma)$ in $\gev^2$
(Sec.~\ref{sec:dimless}). The
critical temperature $T_c$ is defined as the temperature at which the two
competing minima are degenerate,
\begin{align}
	V_{k_{\rm IR}}\bigl(\tilde\rho_{\rm EW}(T_c),0;T_c\bigr)
	=
	V_{k_{\rm IR}}\bigl(0,\tilde\sigma_{\rm s}(T_c);T_c\bigr) ,
\end{align}
where $\tilde\rho_{\rm EW}(T)$ and $\tilde\sigma_{\rm s}(T)$ denote the location of the
electroweak and singlet minima in the dimensionful invariants, tracked continuously as
functions of $T$ from the flowed potential. Because $u_{k_{\rm IR}}$ is obtained
directly from the PINN representation on a continuous domain,
both minima and the saddle point (barrier top) separating
them can be located by a simple gradient-based search on the trained
network, without interpolation artifacts.

Stopping the flow at $k_{\rm IR}=150\gev$ rather than at $k\to0$ is a practical
choice tied to the benchmark ($\Lambda\simeq1108\gev$ over an RG-time range
$t_{\rm range}=-2$; Sec.~\ref{sec:benchmark-point}). Since $k_{\rm IR}\sim T\sim m_h$,
the flow is close to frozen in the outer, deeply broken region, so the residual
running below $k_{\rm IR}$ is a small systematic on the vacuum locations and depths.
It is not a benign one for the barrier, however: continuing to substantially lower
$k_{\rm IR}$ takes the inner, nearly flat region between the minima into the regime
where the potential convexifies and $1+\bar m^2\to0$, where the $\coth$ threshold
functions become singular and a controlled Maxwell-type treatment of the
convexification is required -- the same obstruction behind the floor on
$1+\bar m^2$ and the tachyonic abort of Sec.~\ref{sec:loss}. The inner barrier is
thus both the quantity most sensitive to lowering $k_{\rm IR}$ and the hardest to
resolve, consistent with it being the least converged output of the
two-dimensional network (Sec.~\ref{sec:results-2d}); propagating this systematic
is left for future work.

The strength of the transition is quantified by the standard order
parameter
\begin{align}
	\xi_c \equiv \frac{v_c}{T_c},
	\qquad
	v_c \equiv \sqrt{2\,\tilde\rho_{\rm EW}(T_c)} ,
\end{align}
with $\xi_c \gtrsim 1$ conventionally taken as the (parametric) condition
for a transition strong enough to avoid washout of the baryon asymmetry
in electroweak-baryogenesis scenarios. As in conventional effective-potential
treatments, the LPA$'$ flow here is formulated in a fixed gauge (Landau), so
$\xi_c$ and the barrier inherit the usual gauge dependence of the
effective potential; a manifestly gauge-invariant treatment is beyond the scope
of this proof of concept. Because $T_c$ only signals
degeneracy of the two minima and not the actual completion of the
transition, we additionally estimate the nucleation temperature $T_n<T_c$
from the $O(3)$ bounce action~\cite{Coleman:1977py,Linde:1981zj},
\begin{align}
	S_3(T)
	=
	4\pi \int_0^\infty dr\, r^2
	\left[
	\frac{1}{2}\left(\frac{d\bm\phi}{dr}\right)^2
	+ \Delta V_{k_{\rm IR}}\bigl(\bm\phi(r);T\bigr)
	\right] ,
\end{align}
evaluated on the solution $\bm\phi(r)=(h(r),s(r))$ of the two-field bounce equation
\begin{align}
	\frac{d^2\bm\phi}{dr^2} + \frac{2}{r}\,\frac{d\bm\phi}{dr}
	= \bm\nabla_{\!\bm\phi}\,\Delta V_{k_{\rm IR}}(\bm\phi;T) ,
	\qquad
	\bm\phi'(0)=0, \quad \bm\phi(r\to\infty)\to\bm\phi_{\rm false} ,
\end{align}
where $\bm\phi=(h,s)$ are the canonically normalized radial fields
($\tilde\rho=h^2/2$, $\tilde\sigma=s^2/2$), $\bm\phi_{\rm false}$ is the false
vacuum, and $\Delta V_{k_{\rm IR}}$ is the potential measured relative to the true
vacuum. We solve this for the flowed
two-dimensional potential with the path-deformation method of
Ref.~\cite{Wainwright:2011kj}, which alternates between the one-dimensional
bounce along a fixed field-space path and a deformation of that path toward the
local force. Nucleation is taken to occur when $S_3(T_n)/T_n \simeq 140$
($\simeq 4\ln(M_{\rm Pl}/T)$ at the electroweak scale), the standard criterion
for roughly one bubble per Hubble volume~\cite{Quiros:1999jp,Morrissey:2012db}.
In this work we evaluate $S_3(T)/T$ at a fixed benchmark temperature
directly from the flowed two-dimensional potential
(Sec.~\ref{sec:results-2d}); locating $T_n$, and $T_c$, by a temperature scan
would require retraining the network at each temperature and is left for future
work.

\FloatBarrier
\subsection{Ring (Debye mass) resummation and its FRG counterpart}
\label{sec:ring}

In perturbative treatments of the finite-temperature effective potential,
the appearance of IR-sensitive daisy (ring) diagrams associated with the
zero Matsubara mode requires resummation of thermal (Debye) masses in the
scalar and longitudinal gauge propagators~\cite{Arnold:1992rz,Parwani:1991gq}; without this resummation the
loop expansion in the high-temperature regime breaks down order by order
in the coupling. In our framework, the scalar sector of the Higgs–singlet
system is treated nonperturbatively by means of the functional
renormalization group, and the corresponding resummation of thermal
fluctuations is generated automatically by the Matsubara sum and the
successive integration of momentum shells from $k=\Lambda$ down to
$k\to 0$. Concretely, the bosonic contribution in the flow equation,
\begin{align}
	T\sum_{n\in\mathbb{Z}} \;\longrightarrow\;
	\frac{k}{2\pi}
	\sqrt{1+\bar m^2}\,
	\coth\!\left(\frac{\sqrt{1+\bar m^2}}{2\tau}\right) ,
\end{align}
already includes the zero-mode contribution together with all nonzero
Matsubara modes, and the $k$-dependent masses $\bar m^2$ entering this
expression incorporate the effects of fluctuations integrated out at
higher scales. In this sense the FRG flow for the scalar modes performs
an all-orders resummation of the ring diagrams without requiring a
separate insertion of thermal masses into tree-level propagators.

At the same time, the electroweak gauge bosons and the top quark are not
included dynamically in the FRG flow of $\Gamma_k[H,S]$ but are treated as
an external perturbative sector, entering the scalar flow through the term
$E_{\rm ext}$ of Eq.~\eqref{eq:E-ext}. Inside the flow this term is
evaluated with tree-level (running-coupling) gauge and top masses only,
with no Debye improvement of the longitudinal modes, so its shell
integration reconstructs the one-loop gauge/top contribution to the
potential \emph{without} the daisy resummation. The ring resummation of
this external sector is instead restored a posteriori, by adding it once
and explicitly to the flowed potential -- the ``NN$+$ring'' construction
used for the comparisons of Sec.~\ref{sec:results} -- and it is likewise
built by hand into the perturbative finite-temperature reference curves.
Inserting Debye masses into $E_{\rm ext}$ itself would double-count this
thermal self-energy against that a posteriori correction. Ring resummation
is thus unnecessary within the FRG treatment of the scalar Higgs--singlet
fields, whose flow includes it automatically through the full Wetterich
equation, but it remains essential for the externally supplied gauge and
Yukawa contributions in order to control their high-temperature infrared
behaviour consistently in our hybrid setup.

\FloatBarrier
\section{Benchmark: a grid-based Newton--Krylov relaxation solver}
\label{sec:relax-benchmark}

Before turning to the PINN construction, it is useful to record what happens if the
flow equation derived above is instead attacked with a conventional grid-based
method, both to motivate the mesh-free approach of Sec.~\ref{sec:network} and because the resulting
(non-)convergence pattern is itself physically informative about the equation being
solved.

\FloatBarrier
\subsection{Benchmark point and input parameters}
\label{sec:benchmark-point}

The flow is followed over the RG-time range $t\in[t_{\rm range},0]$, with
$t_{\rm range}=-2$ (the same range used for the multi-block training of
Sec.~\ref{sec:domain-decomp}), corresponding to a physical scale range
\begin{align}
	k_{\rm IR} = 150\gev
	\;\le\;
	k
	\;\le\;
	\Lambda \equiv k_{\rm UV} = k_{\rm IR}\,e^{-t_{\rm range}} \simeq 1108\gev .
\end{align}
The Higgs-sector tree-level parameters entering Eq.~\eqref{eq:u-uv} are not
free: they are fixed by the physical Higgs pole mass and electroweak VEV,
$m_h=125.2\gev$ and $v=246.2\gev$ (i.e.\ $\langle h\rangle=v$,
$\tilde\rho|_{\rm EW}=v^2/2$), through the standard tree-level relations
\begin{align}
	\lambda_H^{\rm tree} = \frac{m_h^2}{2v^2} \simeq 0.129 ,
	\qquad
	\mu_H^{2,\rm tree} = -\lambda_H^{\rm tree}v^2 = -\frac{m_h^2}{2} \simeq -7838\gev^2 .
\end{align}
The singlet sector, by contrast, is unconstrained by any existing
measurement, and we simply choose a tree-level benchmark point
\begin{align}
	\lambda_S^{\rm tree} = 0.27 ,
	\qquad
	\lambda_{HS}^{\rm tree} = 1.0 ,
	\qquad
	\mu_S^{2,\rm tree} = -(90\gev)^2 = -8100\gev^2 ,
\end{align}
with $\lambda_{HS}$ the Higgs--singlet portal coupling.

These tree-level values are then promoted to couplings renormalized at
$\mu=k_{\rm IR}$ by including the leading loop correction. Together with
the one-loop-matched gauge couplings and a top Yukawa
$y_t(k_{\rm IR})\simeq0.9$ (consistent with the $\overline{\rm MS}$ top Yukawa at
this scale), so that
$g_1,g_2,g_3,y_t\simeq0.358,0.649,1.176,0.9$ at $k_{\rm IR}$, we form the
one-loop Coleman--Weinberg potential built from the same tree-level bare
masses as Eq.~\eqref{eq:mass-eigen} (the two scalar mass eigenstates $\bar
m_\pm^2$, $W$, $Z$, and top),
\begin{align}
	u_{\rm CW}(\rho,\sigma)
	=
	\frac{1}{64\pi^2}
	\sum_i n_i\, \bar m_i^4(\rho,\sigma)
	\left[\ln \bar m_i^2(\rho,\sigma) - c_i\right] ,
	\label{eq:u-cw}
\end{align}
with dof/scheme constants $(n_i,c_i)$ equal to $(1,3/2)$ for each of $\bar
m_\pm^2$, $(6,5/6)$ for $W$, $(3,5/6)$ for $Z$, and $(-12,3/2)$ for the top
(the minus sign reflecting Fermi--Dirac statistics), all evaluated at
$\mu=k=k_{\rm IR}$. The Goldstone loop is deliberately left out of this
sum: at the tree-level electroweak point the Goldstone mass $\bar m_G^2$
vanishes exactly [Eq.~\eqref{eq:mass-goldstone} evaluated at
$(\rho,\sigma)=(v^2/2k_{\rm IR}^2,0)$ gives $\bar m_G^2=0$ by construction,
since this is precisely the direction of the spontaneously broken
generator], so its contribution to the \emph{curvature} of $u_{\rm CW}$
(though not its value) hits the well-known Goldstone-boson-catastrophe log
divergence~\cite{EliasMiro:2014pca,Martin:2014bca}: $\partial^2/\partial
m_G^2[\,m_G^2(\ln m_G^2-c)\,]\propto\ln m_G^2\to-\infty$ as $m_G^2\to0$. In
a numerical implementation this divergence is only cut off by whatever ad
hoc infrared regulator is used to keep $\ln|m_G^2|$ finite, and the
resulting matched couplings are highly sensitive to that regulator rather
than converged; the other four loops above are all evaluated at genuinely
nonzero tree masses and are unaffected. We therefore only include the
Goldstone contribution, as usual, in the full finite-temperature potential
used elsewhere in this paper (Sec.~\ref{sec:ewpt}, and the full flow equation solved in
the remainder of this section), where its thermal Debye mass regulates it,
and not in this $T=0$ input-parameter matching step. Evaluating the first
and second derivatives of $u_{\rm CW}$ at the electroweak point
$(\rho,\sigma)=\big(v^2/(2k_{\rm IR}^2),0\big)$ and subtracting them from
the tree-level parameters defines the renormalized couplings at
$k_{\rm IR}$,
\begin{align}
	\lambda_H(k_{\rm IR}) = \lambda_H^{\rm tree} - \tfrac12 \partial_\rho^2 u_{\rm CW}\big|_{\rm EW} ,
	&\qquad
	\mu_H^2(k_{\rm IR}) = \mu_H^{2,\rm tree} - k_{\rm IR}^2\,\partial_\rho u_{\rm CW}\big|_{\rm EW} ,
	\nonumber\\
	\lambda_S(k_{\rm IR}) = \lambda_S^{\rm tree} - \tfrac12 \partial_\sigma^2 u_{\rm CW}\big|_{\rm EW} ,
	&\qquad
	\mu_S^2(k_{\rm IR}) = \mu_S^{2,\rm tree} - k_{\rm IR}^2\,\partial_\sigma u_{\rm CW}\big|_{\rm EW} ,
	\nonumber\\
	\lambda_{HS}(k_{\rm IR}) = \lambda_{HS}^{\rm tree} - \partial_\rho\partial_\sigma u_{\rm CW}\big|_{\rm EW} , &
	\label{eq:cw-matching}
\end{align}
i.e.\ each renormalized parameter is the tree value minus the
corresponding curvature of the one-loop effective potential, so that once
the one-loop piece is added back the renormalized point reproduces the
physical Higgs mass and the chosen singlet benchmark, rather than only at
tree level. The $k_{\rm IR}$ column of Table~\ref{tab:params} lists the
resulting $\lambda_H(k_{\rm IR}),\lambda_S(k_{\rm IR}),
\lambda_{HS}(k_{\rm IR}),\mu_H^2(k_{\rm IR}),\mu_S^2(k_{\rm IR})$. Starting
from this point at $\mu_0=k_{\rm IR}$, the same five couplings, together
with the gauge and top Yukawa couplings, are evolved with the two-loop RGEs
(Appendix~\ref{app:rge}) up to $\mu=\Lambda=k_{\rm UV}$; the resulting
values fix $a_H,a_S,\lambda_H,\lambda_S,\lambda_{HS}$ in the UV tree
potential of Eq.~\eqref{eq:u-uv} and make up the second column of
Table~\ref{tab:params}. The same running couplings, evaluated at
intermediate $t$, also supply the $t$-dependent tree-level reference
entering the analytic seed potential of Eq.~\eqref{eq:useed} and the
perturbative comparison curves used throughout the paper.

\begin{table}[t]
\centering
\small
\begin{tabular}{lcc}
\hline\hline
Parameter & at $k_{\rm IR}=150\gev$ & at $\Lambda=k_{\rm UV}\simeq1108\gev$ \\
\hline
$\lambda_H$      & 0.131  & 0.112 \\
$\lambda_S$      & 0.271  & 0.317 \\
$\lambda_{HS}$   & 1.001  & 1.131 \\
$\mu_H^2$ [$\gev^2$] & $-8334$ & $-8870$ \\
$\mu_S^2$ [$\gev^2$] & $-7917$ & $-8551$ \\
\hline\hline
\end{tabular}
\caption{Higgs--singlet quartic couplings and mass parameters at the IR end
of the flow (after the one-loop Coleman--Weinberg matching described in the
text) and at the UV end (after two-loop RGE running), used to fix
$a_H,a_S,\lambda_H,\lambda_S,\lambda_{HS}$ in the UV boundary potential of
Eq.~\eqref{eq:u-uv}. At $\Lambda$ these correspond to the dimensionless
$a_H=\mu_H^2/\Lambda^2\simeq-0.0072$ and $a_S=\mu_S^2/\Lambda^2\simeq-0.0070$.}
\label{tab:params}
\end{table}

\FloatBarrier
\subsection{Method}
\label{sec:relax-method}

Rather than integrating the flow forward in $t$ from the UV boundary condition by
explicit or adaptive shooting, we discretize scale and field space together and solve
for the entire trajectory $U(t,\rho,\sigma)$ as a single global system: field
derivatives are taken with a second-order central-difference stencil on a uniform
$(\rho,\sigma)$ grid, so that $\mathrm{RHS}(t,U)$ below denotes the right-hand side of
the flow equation~\eqref{eq:flow-dimless} -- already written with $\partial_t u$
isolated on the left -- evaluated as a function of $U$ and its central-difference
field derivatives at fixed $t$ on the grid. The $t$ direction is then discretized with
Crank--Nicolson, so that the residual
\begin{align}
R^n(U^1,\dots,U^{N_t}) &\equiv \left(U^n-U^{n-1}\right)
- \frac{\Delta t}{2}\Bigl[\mathrm{RHS}(t^{n-1},U^{n-1})+\mathrm{RHS}(t^n,U^n)\Bigr] \nonumber\\
&= 0, \qquad n=1,\dots,N_t,
\end{align}
with $U^0$ fixed to the UV tree-level potential $u_\Lambda$ of Eq.~\eqref{eq:u-uv}, is
required to vanish simultaneously at every time slice. In the tree-only limit ($\eta=0$, no
$E_{\rm ext}/{\rm coth}$ term) this system is linear in $U$, so Newton iteration is
unnecessary and the block bidiagonal system is solved exactly by a single LU
factorization; validated against the exact characteristic-curve solution, this scheme
reduces the relative error from $27\%$, obtained with the original upwind/shooting
discretization, to $0.02\%$--$0.6\%$ depending on grid resolution, showing that the
dominant source of error in the shooting approach was the first-order upwind spatial
stencil rather than the marching direction itself.

Once the loop term $E_{\rm ext}$ and the coth threshold functions are switched on the
system becomes nonlinear, and we solve it with a Jacobian-free Newton--Krylov
method~\cite{Knoll:2003nfy}
(the \texttt{newton\_krylov} routine from \texttt{scipy.optimize}, with an
LGMRES~\cite{Baker:2005lgmres} inner solver), seeded by the closed-form sum of the tree solution and a fixed coarse
Debye ansatz for the thermal piece. This is the regime in which a relaxation method
could in principle realize its main advantage over shooting: convergence of the
entire trajectory to a single self-consistent fixed point, rather than the
one-directional accumulation of error characteristic of forward marching.

\FloatBarrier
\subsection{A first check: the Higgs-only reduction (relaxation solver)}
\label{sec:relax-higgs-only}

Before attacking the full two-field problem it is worth checking the solver on the
simplest version of it: the singlet-free reduction of the model -- Higgs plus the
same external gauge and top sector -- in which the radial sector is
one-dimensional. The only scalar masses are then the Goldstone mass
$\bar m_G^2 = u_\rho$ and the single radial mass
$\bar m_H^2 = u_\rho + 2\rho\,u_{\rho\rho}$, the loop sum reducing to
$3/\!\sqrt{1+\bar m_G^2}\,\coth(\cdots) + 1/\!\sqrt{1+\bar m_H^2}\,\coth(\cdots)$,
so that the $2\times2$ radial mass matrix of Eq.~\eqref{eq:mass-matrix}, its
near-degenerate eigenvalue mixing and the associated $\sqrt{\rm disc}$ -- a
natural first suspect for the stall of the full two-field Newton--Krylov
iteration (Sec.~\ref{sec:relax-convergence}) -- are absent by construction. The regulator, the one-loop
Coleman--Weinberg matching, the two-loop running couplings, the
finite-temperature treatment, the central-difference/Crank--Nicolson
discretization and the Newton--Krylov solver are all kept identical.

\begin{figure}[t]
\centering
\includegraphics[width=0.85\linewidth]{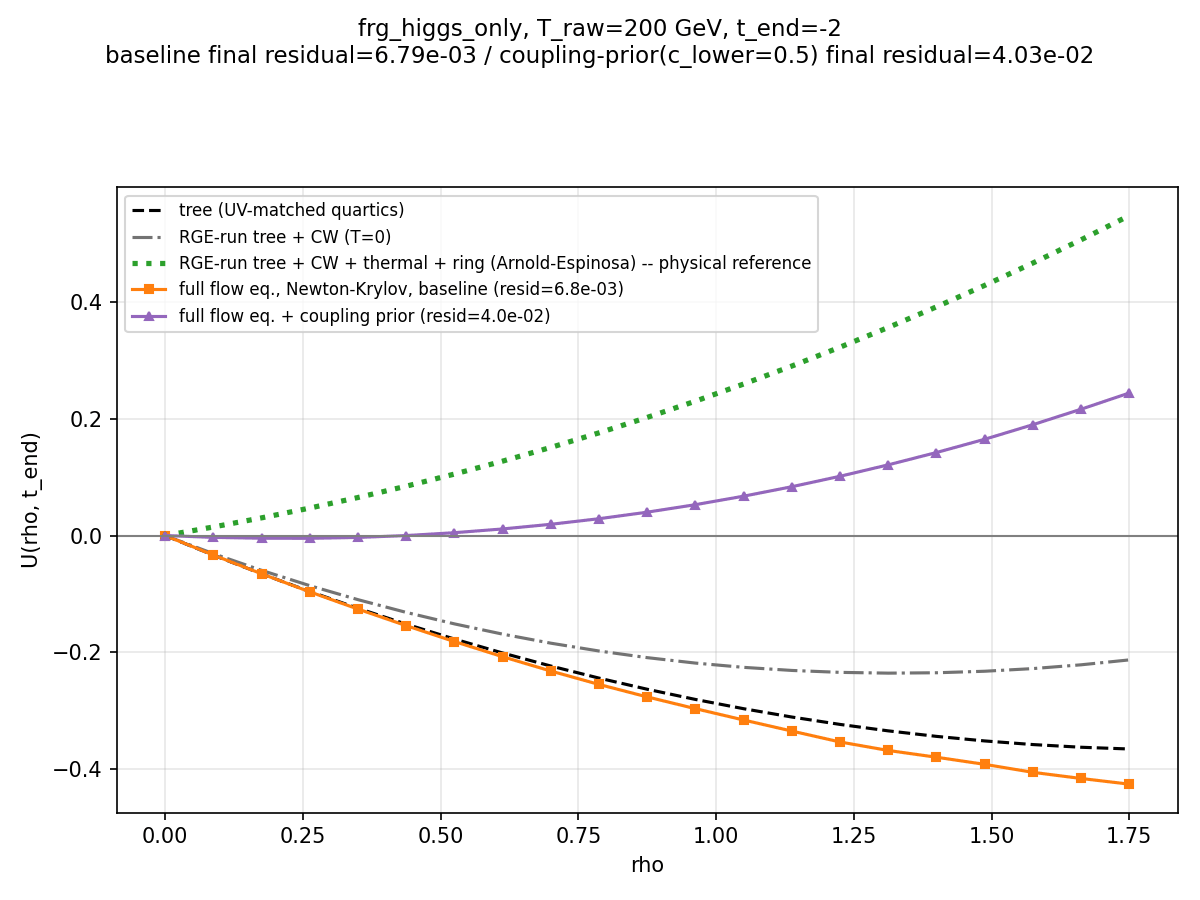}
\caption{First check on the Higgs-only reduction of the grid-based benchmark at
$T=200~\mathrm{GeV}$ and the IR end of the flow $t=t_{\rm range}=-2$. The unforced
Newton--Krylov solution of the full flow equation (orange squares) undershoots the
ring-resummed (Arnold--Espinosa) finite-temperature reference (green dotted, the
closest available analogue of the true nonperturbative trajectory) badly, missing
essentially the entire thermal lift; the tree-level and zero-temperature
Coleman--Weinberg curves (black dashed, grey dash-dotted) are shown for context.
Adding the coupling-prior forcing term with an explicit finite-temperature mass
target (purple triangles) moves the solution toward that reference but still
undershoots it by about a factor of two, while raising the residual norm
$\|R\|_2$ from $6.8\times10^{-3}$ to $4.0\times10^{-2}$. All curves are shifted so
that $u=0$ at $\rho=0$. (At $T=100~\mathrm{GeV}$, not shown, the unforced solver
instead reaches tolerance but converges to a spurious deep well; see the text.)}
\label{fig:relax_higgs_only}
\end{figure}

Removing the singlet does not, however, make the loop-level flow easy to solve
reliably, and the behaviour differs sharply between the two benchmark
temperatures. At $T=100~\mathrm{GeV}$ the unforced baseline Newton--Krylov
iteration \emph{does} reach the requested tolerance $f_{\rm tol}=10^{-8}$
(``converged'' by the solver's own criterion), but the solution it converges to
is unphysical: it decreases monotonically to $u\simeq-0.96$ at
$\rho=\rho_{\max}=1.75$, far below both the tree-level potential ($u=-0.37$ there)
and the ring-resummed (Arnold--Espinosa) finite-temperature reference
($u=-0.06$) -- a spurious deep well rather than the physical potential. A small
residual therefore does not certify a physically correct solution even in this
single-field case. At $T=200~\mathrm{GeV}$ the iteration instead does not converge
at all: the residual norm $\|R\|_2$ stalls at $6.8\times10^{-3}$ (below the
$0.14$--$0.73$ range found for the full two-field system below,
Table~\ref{tab:relax}, though part of that gap simply reflects the smaller
one-dimensional grid), and the stalled solution undershoots the
finite-temperature reference badly ($u=-0.43$ at $\rho_{\max}$ versus $u=+0.55$),
missing essentially the entire thermal lift of the potential
(Fig.~\ref{fig:relax_higgs_only}). Adding the coupling-prior forcing term defined
in Sec.~\ref{sec:coupling-prior} below -- a soft perturbative-consistency penalty
added directly to the residual, here supplemented by an explicit
finite-temperature target on the mass-like combination -- pulls the $T=200$
solution toward the physical reference, raising $u(\rho_{\max})$ to $+0.24$, but
it still undershoots by roughly a factor of two and raises the residual norm to
$4.0\times10^{-2}$.

So neither temperature yields a trustworthy single-field solution: a
converged-but-spurious one at $T=100~\mathrm{GeV}$, an unconverged one at
$T=200~\mathrm{GeV}$. Since the reduction contains none of the singlet
mass-matrix structure, this difficulty cannot be attributed to the
eigenvalue mixing -- it is intrinsic to the nonlinear loop-plus-thermal flow
equation. The full two-field system, to which we now turn, adds the
near-degenerate eigenvalue mixing of Eq.~\eqref{eq:mass-eigen} along the
symmetry axes; we find below that this near-degeneracy is not what obstructs
the solver either. The same Higgs-only reduction is
revisited for the PINN in Sec.~\ref{sec:results-higgs-only}, with a different
outcome.

\FloatBarrier
\subsection{Convergence failure and its origin}
\label{sec:relax-convergence}

Turning to the full two-field system, Newton--Krylov iteration on the full
equation does not converge to the
requested tolerance ($f_{\rm tol}=10^{-8}$)\footnote{\texttt{scipy.optimize.newton\_krylov}'s
convergence criterion is an absolute tolerance on the max-norm
$\|R\|_\infty\equiv\max_i|R_i|$ of the residual vector, with the relative-tolerance and
step-size checks left at their permissive defaults so that this max-norm condition
alone governs convergence here. So $f_{\rm tol}$ is compared against the point-wise
maximum residual quoted below, not the Euclidean norm $\|R\|_2$ of
Table~\ref{tab:relax}.} for any of the regularization choices we
tried, on a $21\times21\times40$ grid: the residual norm drops rapidly during the
first few iterations (by $\gtrsim 95\%$ relative to the initial warm start) and then
stalls at a plateau of order $10^{-1}$, with the point-wise maximum residual settling
around $6\times10^{-3}$--$1\times10^{-1}$ depending on scheme (Table~\ref{tab:relax}).
Backward-Euler time discretization is markedly worse than Crank--Nicolson, with the
residual norm occasionally increasing over the course of the iteration rather than
decreasing monotonically.

The mechanism one would suspect first -- the $\sqrt{\rm disc}$ in the radial mass
eigenvalues $\bar m_\pm^2$ of Eq.~\eqref{eq:mass-eigen}, non-analytic where the two
radial eigenstates become degenerate (${\rm disc}\to0$) -- is not the cause. The
eigenvalue pair enters the loop only through the symmetric combination
$h(\bar m_+^2)+h(\bar m_-^2)$, with $h(x)\equiv(1+x)^{-1/2}\coth(\sqrt{1+x}/2\tau)$,
in which $\sqrt{\rm disc}$ cancels between the two terms; the only genuine
non-analyticity of $h$ is at $1+x=0$, which stopping the flow at $k_{\rm
IR}=150~\mathrm{GeV}$ keeps well outside the domain ($1+\bar m^2\gtrsim0.6$ on the
tree warm start). Consistently, neither a floor on $1+\bar m^2$
(\texttt{mass\_floor}) nor a hard clip of $\sqrt{\rm disc}$ (\texttt{disc\_cut})
improves convergence; the runs reported here take $\texttt{disc\_cut}=0$, and under
the domain restriction below ${\rm disc}$ stays above $\sim2\times10^{-5}$
everywhere in any case.

What limits the solver is instead poor conditioning near the symmetry axes together
with the nonlinearity of the flow: Eq.~\eqref{eq:flow-dimless} is first order in $t$ --
an ordinary initial-value problem in that variable -- but fully nonlinear in the
field derivatives that build the mass matrix $\bar M^2$ of Eq.~\eqref{eq:mass-matrix}:
$u_{\rho\rho},u_{\sigma\sigma},u_{\rho\sigma}$ (together with $u_\rho,u_\sigma$) enter
$\bar M^2$ linearly, but $\bar M^2$'s eigenvalues $\bar m_\pm^2$ then enter the square
root and $\coth$ threshold functions nonlinearly.\footnote{Concretely, the Newton step
linearizes around the current iterate, so the effective coefficient it sees for
$u_{\rho\rho}$ (and likewise $u_{\sigma\sigma},u_{\rho\sigma}$) is
$\partial R/\partial u_{\rho\rho}\propto h'(\bar m_\pm^2)\,
\partial \bar m_\pm^2/\partial \bar M^2\,\partial \bar M^2/\partial u_{\rho\rho}$,
i.e.\ it depends on the current solution estimate rather than being fixed by the
discretization -- unlike, e.g., a linear diffusion equation, where this coefficient
is a constant known in advance.} The stalled residual is largest near $\rho=0$
and $\sigma=0$, where the invariant chart $(\rho,\sigma)=(\tilde h^2/2,\tilde
s^2/2)$ is singular and the one-sided second-derivative stencils and the $\rho\to0$
regularity condition are stiffest; restricting the domain to $\rho,\sigma\ge10^{-3}$
is the single most effective mitigation we tried, appreciably lowering the final
residual norm. Reformulating the system in first-order form --
auxiliary fields $R\equiv u_\rho$, $S\equiv u_\sigma$, so that no second-derivative
stencil is evaluated anywhere -- did not help further: the final residual norm was
no better than the direct formulation, with or without the restriction. No
combination of these measures reaches $f_{\rm tol}$. As discussed in
Sec.~\ref{sec:relax-comparison}, this is because the nonlinear residual has spurious
near-stationary points close to the true trajectory, so a small pointwise residual
does not by itself certify the physically correct solution.

This is the mitigation the comparison of Sec.~\ref{sec:relax-comparison} below
actually uses: all four variants of Table~\ref{tab:relax} are run with the domain
restricted to $\rho,\sigma\ge10^{-3}$ and $\texttt{disc\_cut}=0$, and the slices shown
there are taken along the resulting inner edge of the domain, $\rho=10^{-3}$ and
$\sigma=10^{-3}$, rather than along the excluded axes $\rho=0,\sigma=0$. The
coupling-prior forcing term introduced next is then tested as a second, independent
mitigation on top of this common domain restriction, rather than as an alternative to
it.

\FloatBarrier
\subsection{Coupling-prior forcing term}
\label{sec:coupling-prior}

Two of the four variants compared below (Sec.~\ref{sec:relax-comparison}) add a
\emph{coupling-prior} forcing term directly to the relaxation residual $R^n$ at
every grid point and time slice, implementing a soft perturbative-consistency
constraint (an analogous construction is used as a PINN loss in Sec.~\ref{sec:loss}, there
built from the network's derivative heads instead of finite differences).

At each grid point, local effective quartic couplings and mass parameters are
read off from the current iterate $U$ using the same finite-difference stencil
already used for the flow operator,
\begin{align}
\lambda_H^{\rm grid} \equiv \tfrac12 u_{\rho\rho}, \quad
\lambda_S^{\rm grid} \equiv \tfrac12 u_{\sigma\sigma}, \quad
\lambda_{HS}^{\rm grid} \equiv u_{\rho\sigma}, &
\\[2pt]
a_H^{\rm grid} \equiv u_\rho - u_{\rho\rho}\,\rho - u_{\rho\sigma}\,\sigma, \quad
a_S^{\rm grid} \equiv u_\sigma - u_{\sigma\sigma}\,\sigma - u_{\rho\sigma}\,\rho . &
\end{align}
For each of these five combinations $X\in\{\lambda_H,\lambda_S,\lambda_{HS},a_H,a_S\}$,
the grid-extracted shift away from its UV tree-level value,
$\Delta X_{\rm grid}\equiv X^{\rm grid}-X_{\rm tree}$, is compared against the
corresponding perturbative shift $\Delta X_{\rm pert}(t)$ obtained from the
two-loop running couplings (Appendix~\ref{app:rge}), through two hinge penalties,
\begin{align}
\ell_{\rm sign}[X] &= s_\beta\!\left(-\,\mathrm{sign}(\Delta X_{\rm pert})\,
\frac{\Delta X_{\rm grid}}{|X_{\rm tree}|}\right),
\\[2pt]
\ell_{\rm mag}[X] &= s_\beta\!\left(\frac{c_-\,|\Delta X_{\rm pert}|-|\Delta X_{\rm grid}|}{|X_{\rm tree}|}\right)
+ s_\beta\!\left(\frac{|\Delta X_{\rm grid}|-c_+\,|\Delta X_{\rm pert}|}{|X_{\rm tree}|}\right),
\end{align}
with $(c_-,c_+)=(0.1,3.0)$, where $s_\beta(x)\equiv\beta^{-1}\ln(1+e^{\beta x})$ is a
softplus hinge -- a smooth $C^\infty$ stand-in for $\mathrm{ReLU}$, recovered as
$\beta\to\infty$. Here a smooth hinge is not optional: the kink of a hard
$\mathrm{ReLU}$ at the constraint boundary acts, for the Jacobian-free
Newton--Krylov solve, as exactly the kind of non-smooth forcing that stalls the
iteration (Sec.~\ref{sec:relax-convergence}), and with a fixed nonzero weight the
prior fails to converge at all unless the corner is removed. We use $\beta=10$,
which keeps the transition width within the $\mathcal{O}(0.1)$ natural scale of the
normalized arguments. $\ell_{\rm sign}[X]$ penalizes a grid-extracted shift
with the wrong sign relative to the perturbative one, while $\ell_{\rm mag}[X]$
penalizes a magnitude more than a factor $10$ below or a factor $3$ above the
perturbative shift. Both are switched off by a relative-size mask wherever
$|\Delta X_{\rm pert}|$ is itself negligible compared to $X_{\rm tree}$, and by a
smooth sigmoid mask in $(\rho,\sigma)$ that confines the penalty to a
neighborhood of the field-space origin where perturbation theory is expected to
be reliable (in practice essentially the full $21\times21$ grid used here). At
the finite temperatures considered below, the sign hinge $\ell_{\rm sign}$ is
evaluated for all five combinations, while the magnitude band $\ell_{\rm mag}$
is applied only to the two mass-like combinations $a_H,a_S$; the three quartic
combinations are instead only bounded not to exceed the tree-level value by more
than a factor \texttt{quartic\_ratio\_cap}$=0.99$, since a clean perturbative
target for the curvature of $u$ is not available once nonanalytic thermal
contributions are present.

The weighted sum of these penalties over the five combinations,
\begin{equation}
w_{\rm sign}\sum_X\ell_{\rm sign}[X]+w_{\rm mag}\sum_X\ell_{\rm mag}[X],
\end{equation}
is
evaluated pointwise -- independently at every grid point and time slice, rather
than averaged over sampled collocation points -- and the resulting value itself
is added directly to $\mathrm{RHS}(t^n,U^n)$ at that point, occupying the same
slot as $E_{\rm ext}$ in Eq.~\eqref{eq:flow-dimless} (rather than being
back-propagated through a network as a loss gradient, since here there are no
network weights to update -- $U$ itself is the solution). The forcing term is
added with small weights $(w_{\rm sign},w_{\rm mag})=(0.015,0.006)$: applied
pointwise, it competes directly with the PDE residual at every single grid point
rather than being averaged over a training batch as in the PINN loss, so a much
smaller weight suffices. In short, this lets us test,
within a purely grid-based Newton--Krylov solve, whether the same mechanism
that keeps the PINN solution close in sign and order of magnitude to the
perturbative expectation near the field-space origin can also stabilize (or
at least regularize) the stalled relaxation iteration described above.

\FloatBarrier
\subsection{Comparison at $T=100$ and $T=200~\mathrm{GeV}$}
\label{sec:relax-comparison}

\begin{figure}[t]
\centering
\includegraphics[width=0.99\linewidth]{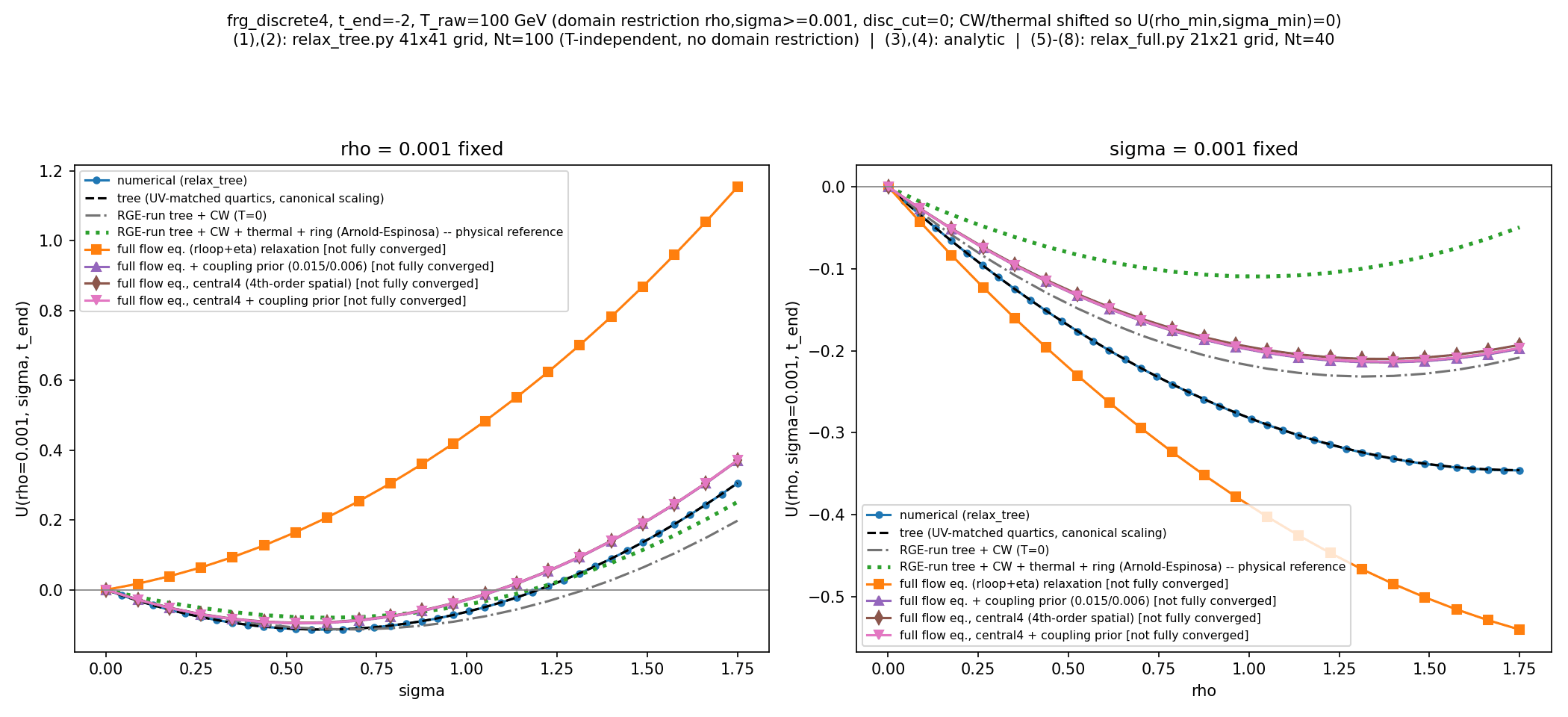}
\caption{Slices of the effective potential $u(t_{\rm range},\rho,\sigma)$ along
$\rho=10^{-3}$ (left) and $\sigma=10^{-3}$ (right) at $T=100~\mathrm{GeV}$, with the
domain restricted to $\rho,\sigma\ge10^{-3}$, comparing the finite-difference
relaxation solution of the tree-only flow (numerical, blue) against its exact
counterpart (tree, black dashed), a static one-loop Coleman--Weinberg reference
(grey dash-dotted), the same reference with the Arnold--Espinosa ring-resummed
finite-temperature one-loop potential added (green dotted, ``physical
reference''), and the four non-converged Newton--Krylov variants of
the full flow equation described in the text (orange, purple, brown, pink). All
curves are shifted so that $u=0$ at $(\rho,\sigma)=(10^{-3},10^{-3})$.}
\label{fig:relax_T100}
\end{figure}

\begin{figure}[t]
\centering
\includegraphics[width=0.99\linewidth]{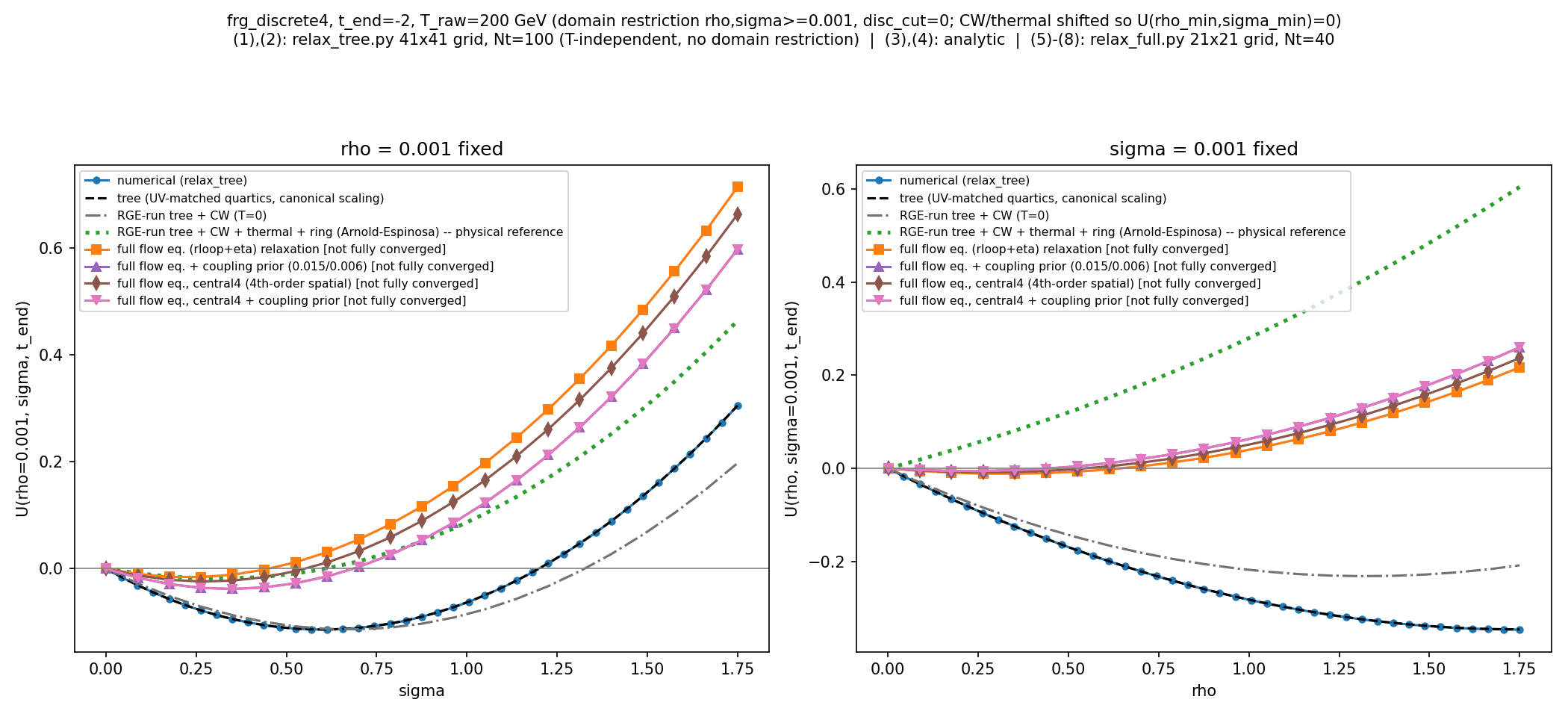}
\caption{Same as Fig.~\ref{fig:relax_T100}, at $T=200~\mathrm{GeV}$. Along
$\rho=10^{-3}$ (left) the coupling-prior and fourth-order-plus-prior variants
(purple, pink) lie almost on top of each other, with the fourth-order variant
without the prior (brown) sitting between them and the unforced baseline (orange);
along $\sigma=10^{-3}$ (right) this separation largely disappears and all four
variants, including the baseline, cluster within $\sim20\%$ of each other while
undershooting the thermally lifted reference curve (green dotted) by roughly a factor
of two.}
\label{fig:relax_T200}
\end{figure}

Figures~\ref{fig:relax_T100} and~\ref{fig:relax_T200} compare, at the IR end of the
flow ($t=t_{\rm range}=-2$) and along $\rho=10^{-3}$ (the singlet direction, with the
Higgs field held near zero) and $\sigma=10^{-3}$ (the Higgs direction), the two inner
edges of the restricted domain, four variants of the (non-converged) Newton--Krylov
solution of the full flow equation, all run with the domain restriction
$\rho,\sigma\ge10^{-3}$ and $\texttt{disc\_cut}=0$ of Sec.~\ref{sec:relax-convergence} -- a baseline run using
the second-order central stencil with no forcing term; the same second-order stencil
with the coupling-prior forcing term of Sec.~\ref{sec:coupling-prior} added, with
weights $(0.015,0.006)$; a fourth-order central spatial stencil with no forcing term;
and the fourth-order stencil combined with the forcing term -- against three
reference curves that do not rely on solving the full equation: the exact tree-level
solution, the tree solution supplemented by a static one-loop Coleman--Weinberg
correction evaluated on the running couplings, and the same supplemented additionally
by the Arnold--Espinosa ring-resummed finite-temperature one-loop potential (the
same physical reference used for the PINN in Sec.~\ref{sec:results}). Final residual
norms and wall-clock times for all eight runs (two temperatures times four variants)
are collected in Table~\ref{tab:relax}.

The comparison illustrates three points. First, the unforced second-order baseline is
a clear outlier at $T=100~\mathrm{GeV}$: it reaches $u=1.15$ at
$(\rho,\sigma)=(10^{-3},\sigma_{\max})$, more than four times the reference value
$u=0.25$ there, while along $\sigma=10^{-3}$ it lands at $u=-0.54$ against a reference
of $u\simeq-0.05$. At $T=200~\mathrm{GeV}$ the baseline is a less extreme outlier:
along $\rho=10^{-3}$ it overshoots the reference only mildly ($u=0.71$ versus
$u=0.46$, a factor $\sim1.5$), while along $\sigma=10^{-3}$ it undershoots by a factor
of $\sim2.8$ ($u=0.22$ versus a reference of $u=0.60$) -- the same direction of
failure as the other three variants there (see below), rather than the
qualitatively different behaviour at $T=100~\mathrm{GeV}$, where the unforced
baseline is a gross outlier from both the reference and the tightly clustered
other three variants (by a factor of three or more). Second, the fate of the other three
variants is temperature- and slice-dependent. At $T=100~\mathrm{GeV}$, all three --
second-order-with-prior, fourth-order-without-prior, and fourth-order-with-prior --
collapse onto essentially one curve regardless of whether the coupling prior is
present, agreeing to within a few percent of each other pointwise
($u(10^{-3},\sigma_{\max})=0.370,0.371,0.371$ and $u(\rho_{\max},10^{-3})=-0.198,
-0.193,-0.196$ for prior, central4, central4+prior respectively). At
$T=200~\mathrm{GeV}$, along $\rho=10^{-3}$ only the two variants that include the
coupling-prior forcing term cluster this tightly
($u(10^{-3},\sigma_{\max})=0.599,0.598$ for prior and central4+prior), while the
fourth-order stencil \emph{without} the prior sits at a markedly different,
intermediate value ($u=0.664$) between the baseline ($u=0.71$) and the prior-forced
pair; along $\sigma=10^{-3}$, by contrast, all four variants including the baseline
are within $\sim20\%$ of each other ($u(\rho_{\max},10^{-3})=0.217,0.260,0.237,0.260$
for baseline, prior, central4, central4+prior respectively), so the sharp
prior-versus-no-prior split seen along $\rho=10^{-3}$ at $T=200~\mathrm{GeV}$ does not
carry over to the other slice. So stencil-order-insensitivity is not a fixed property
of the setup: it holds at $T=100~\mathrm{GeV}$ on both slices, and at
$T=200~\mathrm{GeV}$ only on the $\sigma=10^{-3}$ slice, where it is instead the
presence of the coupling-prior forcing term that separates the curves. Third, where a
tight cluster does form, it is not uniformly close to the finite-temperature
reference. At $T=200~\mathrm{GeV}$ the prior-pair cluster \emph{overshoots} the
reference along $\rho=10^{-3}$ by about $30\%$ ($u\simeq0.60$ versus $u=0.46$) but
\emph{undershoots} it along $\sigma=10^{-3}$ by roughly a factor of two ($u\simeq0.26$
versus $u=0.60$), essentially the same shortfall seen in the baseline there. At
$T=100~\mathrm{GeV}$ the cluster overshoots the reference along $\rho=10^{-3}$ by
about $50\%$ ($u\simeq0.37$ versus $u=0.25$) and sits well below it along
$\sigma=10^{-3}$ ($u\simeq-0.20$ versus a reference of $u\simeq-0.05$). So the
direction of the mismatch (over- or under-shoot) and its
size both vary with slice and temperature, rather than settling into a single clean
pattern.

The true solution of the flow
equation -- the one that actually
satisfies the Wetterich equation together with the correct initial and boundary
conditions -- would, as a consequence of the equation's own physics, automatically be
loosely consistent with the perturbative RGE running near the field-space origin;
nothing needs to be imposed by hand for the exact trajectory to have this property.
The coupling-prior penalty of Sec.~\ref{sec:coupling-prior} was added precisely
because the equation's nonlinearity (the steep, poorly conditioned regime discussed
above) lets the Newton--Krylov iterate drift away from that true trajectory while still keeping
the pointwise residual $R$ small, so a small $\|R\|$ alone cannot certify that an
iterate is still on-trajectory; loose perturbative consistency was meant to serve as
an independent, cheap check on this drift, and the forcing term as a way to pull the
iteration back when it is violated. Even so, Table~\ref{tab:relax} shows this check
mostly failing to close: turning the prior on \emph{increases} $\|R\|$ in three of
the four baseline/central4 pairs (from $0.144$ to $0.217$ and from $0.182$ to $0.220$
at $T=100~\mathrm{GeV}$; from $0.256$ to $0.292$ at $T=200~\mathrm{GeV}$ for the
central4 pair) rather than driving it toward zero. The one exception is the
second-order pair at $T=200~\mathrm{GeV}$, where adding the prior actually
\emph{lowers} $\|R\|$ (from $0.728$ to $0.287$) -- but this is also the one case
where the unforced baseline itself is the worst-converged of all eight runs, so the
improvement reads less as the prior successfully enforcing consistency and more as a
regression toward the typical $\|R\|\sim0.2$--$0.3$ plateau seen elsewhere in the
table. Either way, none of the eight runs comes close to the requested tolerance, and
the penalty itself is not being satisfied at the stalled solution. In short, even this additional
mechanism does not bring the drift under control, which is consistent with, and gives
a mechanistic reason for, the slice- and temperature-dependent mismatch noted above. Taken together,
these observations indicate that the addition of an external forcing term, rather
than genuine convergence of the Newton--Krylov iteration, is what determines the fate
of the stalled solution, which is the opposite of what a reliable solver should
exhibit and is the practical motivation for the PINN-based construction of Sec.~\ref{sec:network}.

\begin{table}[t]
\centering
\small
\begin{tabular}{lcccc}
\hline\hline
Variant & \multicolumn{2}{c}{$T=100~\mathrm{GeV}$} & \multicolumn{2}{c}{$T=200~\mathrm{GeV}$} \\
 & final $\|R\|$ & wall time [s] & final $\|R\|$ & wall time [s] \\
\hline
baseline & 0.144 & 778 & 0.728 & 768 \\
+ coupling prior & 0.217 & 609 & 0.287 & 627 \\
central4 & 0.182 & 2026 & 0.256 & 1974 \\
central4 + coupling prior & 0.220 & 1396 & 0.292 & 1395 \\
\hline\hline
\end{tabular}
\caption{Final residual norm $\|R\|_2$ and wall-clock time of the Jacobian-free
Newton--Krylov relaxation on a $21\times21$ spatial grid with $N_t=40$ Crank--Nicolson
steps, for the four variants shown in
Figs.~\ref{fig:relax_T100}--\ref{fig:relax_T200}. All four use the domain restriction
$\rho,\sigma\ge10^{-3}$ of Sec.~\ref{sec:relax-convergence} with $\texttt{disc\_cut}=0$; the coupling-prior
variants additionally add the forcing term of Sec.~\ref{sec:coupling-prior} with
weights $(0.015,0.006)$. All runs use $f_{\rm tol}=10^{-8}$ and a maximum of 100
outer Newton iterations. None of the eight runs reaches the requested tolerance.}
\label{tab:relax}
\end{table}

\FloatBarrier
\section{Network structure and loss functions}
\label{sec:network}

\FloatBarrier
\subsection{Network}
\label{sec:network-arch}
The full potential is decomposed, in the spirit of the physics-motivated trial
solutions of Ref.~\cite{Lagaris:1997bcy}, into a physics-motivated baseline and a learnable residual:
\begin{equation}
u(t,\rho,\sigma) = \tilde{u}_{\mathrm{tree}}(t,\rho,\sigma) + \tilde{u}_{\mathrm{th}}(t,\rho,\sigma) + N(t,\rho,\sigma).
\end{equation}
Here $\tilde u_{\mathrm{tree}}$ is the running tree-level potential $u_{\Lambda}|_{k_{\rm UV}\to k}$ with each of its five terms rescaled by a fixed $\mathcal{O}(1)$ factor (Appendix~\ref{app:impl}), and $\tilde{u}_{\mathrm{th}} = c_{T,\rho}\,\rho\,\tau^2 + c_{T,\sigma}\,\sigma\,\tau^2$ with $(c_{T,\rho},c_{T,\sigma})=(0.2,0.1)$ is a crude leading high-temperature mass term. These two pieces are not meant to be accurate; they only absorb the bulk of the potential so that the network $N$ is left to represent the smaller, nontrivial part of the flow. This is distinct from the analytic seed $u_{\rm seed}$ of Eq.~\eqref{eq:useed}, which is a genuine one-loop (ring-resummed) thermal potential used only as the fitting target for the network's derivatives at the UV boundary $t=0$.

\begin{figure}[t]
\centering
    \includegraphics[width=0.99\linewidth]{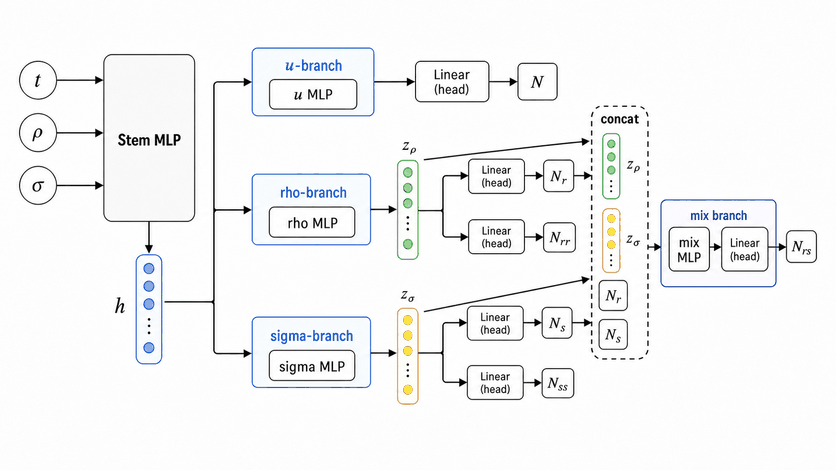}
\caption{ %
Schematic structure of the neural-network ansatz (BaseNet) used to represent the correction to the effective potential and its field derivatives. In the figure the subscripts $\rho,\sigma$ are abbreviated $r,s$ and the branch latent vectors are labelled $z_\rho,z_\sigma$. The three inputs $(t,\rho,\sigma)$ are first processed by a shared stem MLP into a latent representation $h$. From this stem output, three branches are constructed. The $u$-branch applies an additional MLP to $h$ and a linear head to produce the scalar output $N$, which corrects the potential itself. The $\rho$-branch takes the same stem output $h$, passes it through a dedicated MLP to obtain a latent vector $z_\rho$, and then uses two linear heads acting on $z_\rho$ to generate the $\rho$-direction outputs $N_\rho$ and $N_{\rho\rho}$. Similarly, the $\sigma$-branch processes $h$ through another MLP to obtain $z_\sigma$, from which two linear heads produce $N_\sigma$ and $N_{\sigma\sigma}$. The mixed branch receives as input the concatenation of the latent vectors $z_\rho$ and $z_\sigma$ together with the first-derivative outputs $N_\rho$ and $N_\sigma$, but not the second derivatives $N_{\rho\rho}$ or $N_{\sigma\sigma}$. This concatenated vector is passed through a final MLP and a linear head to yield the mixed second derivative $N_{\rho\sigma}$. Thus, the network outputs $(N, N_\rho, N_\sigma, N_{\rho\rho}, N_{\sigma\sigma}, N_{\rho\sigma})$, with the mixed derivative $N_{\rho\sigma}$ explicitly constructed from the field-direction latent features and first derivatives.
}
\label{fig:mynet}
\end{figure}

The neural network $N$ is implemented via a modular architecture (Fig.~\ref{fig:mynet}) that explicitly outputs not only the potential correction but also its first and second derivatives:
\begin{equation}
\left(N, N_\rho, N_\sigma, N_{\rho\rho}, N_{\sigma\sigma}, N_{\rho\sigma}\right).
\end{equation}
We arrived at this design after an unsuccessful attempt at the more direct alternative: obtaining $N_\rho,N_\sigma,N_{\rho\rho},N_{\sigma\sigma},N_{\rho\sigma}$ by nested automatic differentiation of a single-output network with respect to $\rho,\sigma$. Even combined with the same soft consistency penalty used throughout this paper (Sec.~\ref{sec:loss}), this alternative still failed to converge to a physically sensible branch, in the same manner as the failure modes discussed in Sec.~\ref{sec:results-limitations}. Backpropagating the loss gradient through such a network requires third-order mixed field-and-weight derivatives -- the loss depends on a second field derivative of the network, and training differentiates that dependence once more with respect to the weights -- which measurably deepens the computational graph; we cannot, however, cleanly disentangle this added complexity from the same branch-selection non-uniqueness of the flow equation itself, since both were present in this comparison. This design avoids that repeated automatic differentiation and improves computational efficiency, at the cost of no longer guaranteeing that the derivative heads coincide with the true derivatives of $N$ -- addressed only approximately, via the self-consistency penalty of Sec.~\ref{sec:loss} below.

The architecture consists of three components. First, a shared ``stem'' multilayer perceptron (MLP) processes the input $(t,\rho,\sigma)$ into a latent representation, enabling efficient encoding of correlations among scale and field variables.

Second, separate branch networks are introduced for the $\rho$- and $\sigma$-directions. Each branch produces features that are mapped to first- and second-order derivatives ($N_\rho, N_{\rho\rho}$ and $N_\sigma, N_{\sigma\sigma}$, respectively), reflecting the anisotropic role of the field directions in the RG flow.

Third, a mixed branch combines the latent representations from both field directions together with the first derivatives $(N_\rho, N_\sigma)$ to produce the mixed second derivative $N_{\rho\sigma}$. This construction allows the model to capture nontrivial correlations between $\rho$ and $\sigma$.

All output heads are initialized to zero, ensuring that the initial network reproduces the baseline potential. Training then corresponds to learning deviations from the known tree-level and thermal structures, which stabilizes optimization and improves physical interpretability.

To enhance numerical performance, all inputs are linearly rescaled to the interval $[-1,1]$. The corresponding Jacobian factors are consistently incorporated when relating derivatives in scaled and physical coordinates.

Concretely, the stem is a three-layer MLP of width $160$; the $u$-branch is a
single hidden layer of width $160$, the $\rho$- and $\sigma$-branches two hidden
layers of width $160$ each, and the mixed branch two hidden layers of width $320$
(its input being the concatenation $z_\rho\oplus z_\sigma\oplus N_\rho\oplus N_\sigma$,
of dimension $322$). All activations are SiLU and all output heads are
zero-initialized linear maps, giving $\approx 3.9\times10^{5}$ parameters per
\texttt{BaseNet}. One such network is trained per domain block (Sec.~\ref{sec:domain-decomp}),
for a total of $10$ networks along a one-dimensional slice and $50$ for the full
two-field problem. The remaining architectural and training hyperparameters are
collected in Appendix~\ref{app:impl}.

This hybrid construction, combining analytic physics priors with a structured neural network ansatz, provides a flexible yet controlled representation of the effective potential suitable for functional renormalization group studies.

\FloatBarrier
\subsection{Loss functions}
\label{sec:loss}

Collocation points $(t,\rho,\sigma)$ within a given block are resampled at every training iteration from a mixture of $40\%$ points drawn uniformly in the rescaled coordinates $[-1,1]^3$ and $60\%$ points drawn as $\cos\!\left(\pi\,\mathcal{U}(0,1)\right)$, which asymptotically reproduces the clustering of Chebyshev nodes near the block boundaries $\pm1$ and improves the resolution of the fit where the potential and its derivatives vary most rapidly. The network parameters are optimized by minimizing a weighted sum,
\begin{equation}
\mathcal{L}
=
w_{\rm pde}\,\mathcal{L}_{\rm pde}
+ w_{\rm consist}\,\mathcal{L}_{\rm consist}
+ w_{\rm bc}\,\mathcal{L}_{\rm bc}
+ w_{\rm overlap}\,\mathcal{L}_{\rm overlap}
+ w_{\rm sign}\,\mathcal{L}_{\rm sign}
+ w_{\rm mag}\,\mathcal{L}_{\rm mag} ,
\label{eq:loss-total}
\end{equation}
whose six terms we describe in turn.

\paragraph{PDE residual.}
Let $R(t,\rho,\sigma)$ denote the residual obtained by moving every term of the flow equation to one side. Only the RG-time derivative $\partial_t u$ is evaluated by automatic differentiation of the scalar output; all field derivatives entering $R$ (and hence the Goldstone and radial mass eigenvalues $\bar m_G^2,\bar m_\pm^2$) are taken directly from the corresponding output heads described in Sec.~\ref{sec:network-arch}, avoiding repeated automatic differentiation through the field directions. The PDE loss is then
\begin{equation}
\mathcal{L}_{\rm pde} = \left\langle R(t,\rho,\sigma)^2 \right\rangle_{(t,\rho,\sigma)} .
\end{equation}
For numerical stability, $1+\bar m_G^2$ and $1+\bar m_\pm^2$ are floored at $10^{-10}$ before entering the threshold functions, the discriminant ${\rm disc}=(\bar M_{11}^2-\bar M_{22}^2)^2+4(\bar M_{12}^2)^2$ of the radial mass matrix [Eq.~\eqref{eq:mass-matrix}] is floored at $10^{-6}$ (so $\sqrt{\rm disc}\ge10^{-3}$), and the resulting loop sum is clamped to $[-2,2]$. The comparatively large \mbox{$\mathrm{disc}$-floor} is a training-stability measure rather than a physical regulator: although $\sqrt{\rm disc}$ cancels from the loop sum itself (the two radial eigenvalues enter only through a combination symmetric under $\bar m_+^2\leftrightarrow\bar m_-^2$, cf.\ Sec.~\ref{sec:relax-convergence}), its gradient is back-propagated through the two eigenvalues along separate paths, and the near-degenerate limit ${\rm disc}\to0$ makes that gradient stiff. Occurrence of $1+\bar m_G^2<0$ or $1+\bar m_\pm^2<0$ after a majority of training has elapsed signals a pathological (tachyonic) configuration and aborts the run, which in practice never triggers once the boundary and consistency losses below have been imposed.

\paragraph{Consistency loss.}
Since $u_\rho,u_\sigma,u_{\rho\rho},u_{\sigma\sigma},u_{\rho\sigma}$ are produced by dedicated output heads rather than obtained by differentiating $u$ itself, nothing in the architecture guarantees that they coincide with the actual derivatives of the scalar output. We therefore add a self-consistency penalty comparing each head to a single automatic differentiation of the lower-order heads,
\begin{align}
\mathcal{L}_{\rm consist}
={}&
\left\langle \left(\partial_\rho u - u_\rho\right)^2 + w_{s1}\left(\partial_\sigma u - u_\sigma\right)^2 \right\rangle
\nonumber\\
&+ \left\langle \left(\partial_\rho u_\rho - u_{\rho\rho}\right)^2
+ w_{s2}\left(\partial_\sigma u_\sigma - u_{\sigma\sigma}\right)^2
+ w_{rs}\Big[\left(\partial_\rho u_\sigma - u_{\rho\sigma}\right)^2+\left(\partial_\sigma u_\rho - u_{\rho\sigma}\right)^2\Big] \right\rangle ,
\end{align}
evaluated at the same collocation points as $\mathcal{L}_{\rm pde}$. The weights $w_{s1}=w_{s2}=w_{rs}=5$ used in the finite-temperature runs reported here upweight the singlet-direction and mixed terms relative to the Higgs-direction ones, whose gradients are typically larger simply because the Higgs direction carries the dominant tree-level curvature.

\paragraph{Boundary and inter-block continuity in $t$.}
At the UV edge $t=0$ of the outermost block in scale, the first derivatives of the network are fit to those of an analytic seed,
\begin{equation}
u_{\rm seed}(t,\rho,\sigma) = u_{\rm tree}(t,\rho,\sigma) + c_{\rm cw}\,u_{\rm CW}(t,\rho,\sigma) + \mathbb{1}_{\rm finite\,T}\,u_{\rm th}^{\rm 1-loop}(t,\rho,\sigma) ,
\label{eq:useed}
\end{equation}
with $c_{\rm cw}$ a switch for the one-loop Coleman--Weinberg piece (set to zero in the runs reported here, so that the network learns any zero-temperature loop correction purely from the flow equation itself) and $u_{\rm th}^{\rm 1-loop}$ the finite-temperature one-loop potential built from the same field content as the flow equation, evaluated with Debye-improved (ring-resummed) masses as discussed in Sec.~\ref{sec:ring}. This gives $\mathcal{L}_{\rm bc}=\langle(\partial_\rho u_{\rm seed}-u_\rho)^2+(\partial_\sigma u_{\rm seed}-u_\sigma)^2\rangle_{t=0}$. For a block that is not the outermost one in $t$, this analytic anchor is replaced by matching against the already-trained, frozen neighboring block $u^{\rm prev}$ evaluated at the shared time slice,
\begin{equation}
\mathcal{L}_{\rm bc} = \left\langle (u_\rho - u_\rho^{\rm prev})^2 + (u_\sigma-u_\sigma^{\rm prev})^2 \right\rangle
+ w_{\rm bc,2}\left\langle (u_{\rho\rho}-u_{\rho\rho}^{\rm prev})^2 + (u_{\sigma\sigma}-u_{\sigma\sigma}^{\rm prev})^2 + (u_{\rho\sigma}-u_{\rho\sigma}^{\rm prev})^2 \right\rangle ,
\end{equation}
enforcing (at least) $C^1$ continuity of the flowed potential across the interface; the weight of the second-derivative term is set to zero in the present runs.

\paragraph{Field-space interface matching.}
The same construction is applied across the boundaries between adjacent field-space blocks (Sec.~\ref{sec:domain-decomp}): for two networks trained on overlapping $\rho$ (and, in the two-dimensional setup, $\sigma$) windows sharing a thin strip of width $\delta\rho=0.02$, the first derivatives are required to agree throughout that strip,
\begin{equation}
\mathcal{L}_{\rm overlap} = \left\langle (u_\rho^{\rm left} - u_\rho)^2 + (u_\sigma^{\rm left} - u_\sigma)^2 \right\rangle_{(t,\rho,\sigma)\,\in\,\text{overlap strip}} ,
\end{equation}
where ``left'' denotes the previously trained, frozen neighboring block (or, in the full 2D grid, the frozen left and bottom neighbors in $\rho$ and $\sigma$, respectively). This is the mechanism referred to in the Introduction as the derivative-matching condition tying the multi-domain decomposition together, in the spirit of the interface conditions of extended PINNs~\cite{Jagtap:2020}.

\paragraph{Soft perturbative consistency (sign and magnitude penalties).}
The remaining two terms implement the soft consistency constraint mentioned in the abstract. On collocation points spanning the field-space domain, at the infrared edge of each RG-time block ($t=-1$ for the UV block, $t=-2$ for the IR block), we extract effective quartic couplings and quadratic mass terms directly from the network's second- and first-derivative heads,
\begin{align}
\lambda_H^{\rm NN} \equiv \tfrac12 u_{\rho\rho}, \quad
\lambda_S^{\rm NN} \equiv \tfrac12 u_{\sigma\sigma}, \quad
\lambda_{HS}^{\rm NN} \equiv u_{\rho\sigma}, &
\\
a_H^{\rm NN} \equiv u_\rho - u_{\rho\rho}\,\rho - u_{\rho\sigma}\,\sigma, \quad
a_S^{\rm NN} \equiv u_\sigma - u_{\sigma\sigma}\,\sigma - u_{\rho\sigma}\,\rho . &
\end{align}
For each of these five combinations $X\in\{\lambda_H,\lambda_S,\lambda_{HS},a_H,a_S\}$ we compare the network's shift away from its UV tree-level value, $\Delta X_{\rm NN}\equiv X^{\rm NN}-X_{\rm tree}$, against the corresponding perturbative shift $\Delta X_{\rm pert}(t)$ obtained from the two-loop running couplings (Appendix~\ref{app:rge}), through two hinge-type penalties,
\begin{align}
\ell_{\rm sign}[X] &= s_\beta\!\left(-\,\mathrm{sign}(\Delta X_{\rm pert})\,\frac{\Delta X_{\rm NN}}{|X_{\rm tree}|}+\text{margin}\right),
\\[2pt]
\ell_{\rm mag}[X] &= s_\beta\!\left(\frac{c_-\,|\Delta X_{\rm pert}|-|\Delta X_{\rm NN}|}{|X_{\rm tree}|}\right)
+ s_\beta\!\left(\frac{|\Delta X_{\rm NN}|-c_+\,|\Delta X_{\rm pert}|}{|X_{\rm tree}|}\right),
\end{align}
with $(c_-,c_+)=(0.1,3.0)$ and $s_\beta$ the same softplus hinge used for the
grid-based prior in Sec.~\ref{sec:coupling-prior}, here with the sharper
$\beta=100$. As in the grid case a smooth hinge rather than a hard $\mathrm{ReLU}$
is used deliberately: the $\mathrm{ReLU}$ corner at the constraint boundary was found
to correlate with non-smooth, sign-oscillating behaviour in the learned curvature
$u_{\rho\rho}$, and keeping the penalty differentiable across its activation threshold
gives visibly cleaner second derivatives. $\mathcal{L}_{\rm sign}$ and $\mathcal{L}_{\rm mag}$ are then obtained by averaging $\ell_{\rm sign}[X]$ and $\ell_{\rm mag}[X]$ over $X$ with relative weights of order unity (an order of magnitude smaller for $\lambda_{HS}$, whose perturbative shift is comparatively less reliable), restricted by a further relative-size mask that switches the penalty off wherever $|\Delta X_{\rm pert}|$ itself is negligible compared to $X_{\rm tree}$. The magnitude window is intentionally loose: rather than forcing exact agreement with the perturbative result, which would defeat the purpose of a nonperturbative treatment, it merely forbids the trained network from developing quartic or mass corrections with the wrong sign or a wildly different order of magnitude from the (reliable, weak-coupling) perturbative expectation. In the finite-temperature runs the penalty is split by a smooth sigmoid mask at $\rho=\rho_{\rm cw}$, $\sigma=\sigma_{\rm cw}$ (Appendix~\ref{app:impl}) into an inner and an outer field-space region. In the inner region, near the axes, the sign hinge is applied only to the singlet mass combination $a_S$ and the magnitude band to $a_S$ (and, at $T=100\gev$, to $a_H$), in each case with the analytic thermal shift of the mass combination subtracted first; the quartics there carry only the flat cap $|\Delta\lambda_i^{\rm NN}/\lambda_{i,\rm tree}|\lesssim 1$, since the nonanalytic thermal contributions near $\rho=0$ preclude a clean perturbative target for the curvature of $u$. In the outer region, where the thermal potential is smooth, the singlet quartic (and, at $T=100\gev$, the Higgs quartic) instead carries a magnitude band around its two-loop RGE-running value. The Higgs mass combination $a_H$ carries no sign constraint at finite temperature. This finite-temperature form is thus considerably more reduced than the grid-based coupling prior of Sec.~\ref{sec:coupling-prior} to which it is otherwise analogous: the grid prior applies the sign hinge to all five combinations and uses the softer $\beta=10$, whereas here only $a_S$ is sign-constrained and $\beta=100$. The full set of weight values is collected in Appendix~\ref{app:impl}.

Soft anchoring of a learned solution to a cheap analytic model of this kind is used elsewhere in scientific machine learning -- as multi-fidelity training of a correction network~\cite{Meng:2020mfd}, and, closest in spirit, as the enforcement of known exact conditions on machine-learned density functionals~\cite{Kirkpatrick:2021dm21}. We arrived at the present construction independently; the difference is that in those settings the analytic guidance refines a model that already works, whereas here, as we discuss next, it turns out to be needed to select the physical branch at all.

In practice this constraint is not merely a convenience: switching it off (setting $w_{\rm sign}=w_{\rm mag}=0$ in Eq.~\eqref{eq:loss-total}) leaves the PDE, consistency, and boundary/interface losses alone unable to reliably pick out a physically sensible solution, since the flow equation's nonlinearity admits nearby stationary points of that reduced loss with quartic or mass corrections of the wrong sign, or an unphysical order of magnitude, relative to the perturbative expectation near the field-space origin. Reaching the results reported below also required hand-tuning the relative weights of the combinations entering $\mathcal{L}_{\rm sign}$ and $\mathcal{L}_{\rm mag}$ (Appendix~\ref{app:impl}), and these weights were not identical between the two benchmark temperatures; we did not find a principled, weight-free way to impose this constraint. We report this openly as a practical limitation of the present setup rather than treat it as incidental, and return to it in Sec.~\ref{sec:results}.

\FloatBarrier
\subsection{Domain decomposition and multi-block training}
\label{sec:domain-decomp}

A single network of the size described in Sec.~\ref{sec:network-arch} cannot accurately represent the potential over the entire physically relevant range. The shape of $u(t,\rho,\sigma)$ changes qualitatively between the UV, where it is close to the tree-level double expansion, and the IR, where it may develop a nontrivial multi-well structure; and fully-connected networks are additionally subject to \emph{spectral bias}, fitting the low-frequency, large-scale part of a target function far faster under gradient descent than its fine-scale features~\cite{Rahaman:2019spectral,Xu:2020fprinciple}. In the PINN setting the same bias acts on the PDE residual and can be read off from the eigenspectrum of the associated neural tangent kernel~\cite{Wang:2022ntk}; it worsens as the range of scales a single network must resolve grows -- here, the nearly flat outer potential together with the sharp curvature that builds up near the emerging minima over the course of the flow. Restricting each subnetwork to a small sub-domain narrows this range and is a known mitigation~\cite{Moseley:2023fbpinn}. We therefore partition the domain into a set of blocks and train one network per block, stitching the pieces together with the boundary and interface losses introduced above -- an approach conceptually similar to the domain decomposition of extended PINNs~\cite{Jagtap:2020}, applied here simultaneously in RG scale and in field space.

In the $t$ direction, the full range $t\in[t_{\rm range},0]$ (with $t_{\rm range}=-2$ in the runs reported here) is split into two contiguous blocks, $t\in[-1,0]$ and $t\in[-2,-1]$, trained sequentially from the UV block to the IR block. The UV block is anchored to the analytic seed potential of Eq.~\eqref{eq:useed} at $t=0$, while the IR block is instead matched, via $\mathcal{L}_{\rm bc}$, to the already-trained UV block at the shared slice $t=-1$.

In the field directions, $\rho\in[0,\rho_{\max}]$ (or $\sigma\in[0,\sigma_{\max}]$ for the $\sigma$-mode runs, and both for the two-dimensional runs) with $\rho_{\max}=\sigma_{\max}=1.75$ in units of $k^2$ is tiled by $n_{\rm blocks}=5$ windows on a uniform spacing $\Delta\rho=0.35$; each window is widened by a thin band $\delta\rho=0.02$ so that it overlaps its right neighbor, and the first derivatives are matched across that overlap by $\mathcal{L}_{\rm overlap}$. In the one-dimensional $\rho$- and $\sigma$-mode runs the network remains a genuine function $u(t,\rho,\sigma)$ of both fields: only the tiled direction is not fixed to a point, while the other one is narrowed to $[0,0.35]$ -- wide enough to give the network real dependence on both fields, and hence a nontrivial mixed derivative $u_{\rho\sigma}$ and consistency loss, without the cost of also tiling it into blocks. Critically, this still solves the same full two-field flow equation on a genuinely two-dimensional domain -- both radial mass eigenvalues, the mixing term $u_{\rho\sigma}$, and the network's mixed-derivative head are all active -- so the $\rho$- and $\sigma$-mode runs are not a simplified single-field problem; only the domain and the block tiling are restricted. The blocks are trained sequentially, each one referencing its already-trained, frozen left neighbor in $\rho$ (and, for the full 2D grid, also its bottom neighbor in $\sigma$, the two field directions being swept as an outer/inner loop over a $5\times5$ grid of windows). Since every field-space window is itself further divided into the two $t$-sub-blocks described above, the potential over the full physically relevant domain is ultimately represented by a patchwork of up to $5\times5\times2=50$ independently trained \texttt{BaseNet} instances for the two-field problem (or $5\times2=10$ along a one-dimensional slice), each one accurate only within, and continuously matched across, its own sub-domain.

\FloatBarrier
\subsection{Optimizer and training schedule}
\label{sec:optimizer}

Each block is trained with the SOAP optimizer~\cite{SOAP}, which combines an Adam-like update rule~\cite{Kingma:2014adam} with a Shampoo-style preconditioner~\cite{Gupta:2018shampoo}, together with a cosine-annealed learning-rate schedule~\cite{Loshchilov:2016sgdr} decaying from a base learning rate $3\times10^{-4}$ to $\eta_{\min}=10^{-5}$ over $10^{4}$ iterations per block. At each iteration the losses are evaluated on freshly resampled collocation points -- $7168$ for the PDE residual, $4096$ for the boundary condition, $2048$ for the field-space interface, and $1024$ for the perturbative-consistency penalty -- rather than on a static grid. Once the main schedule completes, additional short refinement rounds of $100$ iterations each are carried out at the final, frozen learning rate -- without further decay and without intermediate checkpointing -- for up to $25$ such rounds, stopping early as soon as the total loss falls below a fixed threshold ($10^{-9}$). Between successive rounds the acceptance band of the magnitude penalty $\mathcal{L}_{\rm mag}$ is widened by a fixed geometric factor, so that this deliberately loose constraint does not obstruct the final convergence of the PDE and boundary terms. This low-noise fine-tuning stage improves the accuracy of the derivative outputs that the downstream analysis of Sec.~\ref{sec:ewpt} (vacuum locations, bounce action) relies on, without reintroducing the risk of instability that a further scheduled decay of the learning rate could otherwise carry.

Training uses PyTorch on a single NVIDIA RTX A6000 GPU; at $\approx0.2\,{\rm s}$ per
iteration, a full $\rho$- or $\sigma$-direction run ($10$ blocks) takes up to
$\approx7\,{\rm h}$ including refinement rounds, and the two-dimensional run ($50$
blocks) up to $\approx37\,{\rm h}$ -- consistent with the checkpoint/auto-resume design
needed to span multiple Paperspace sessions.

\begin{figure}[t]
\centering
\includegraphics[width=0.75\linewidth]{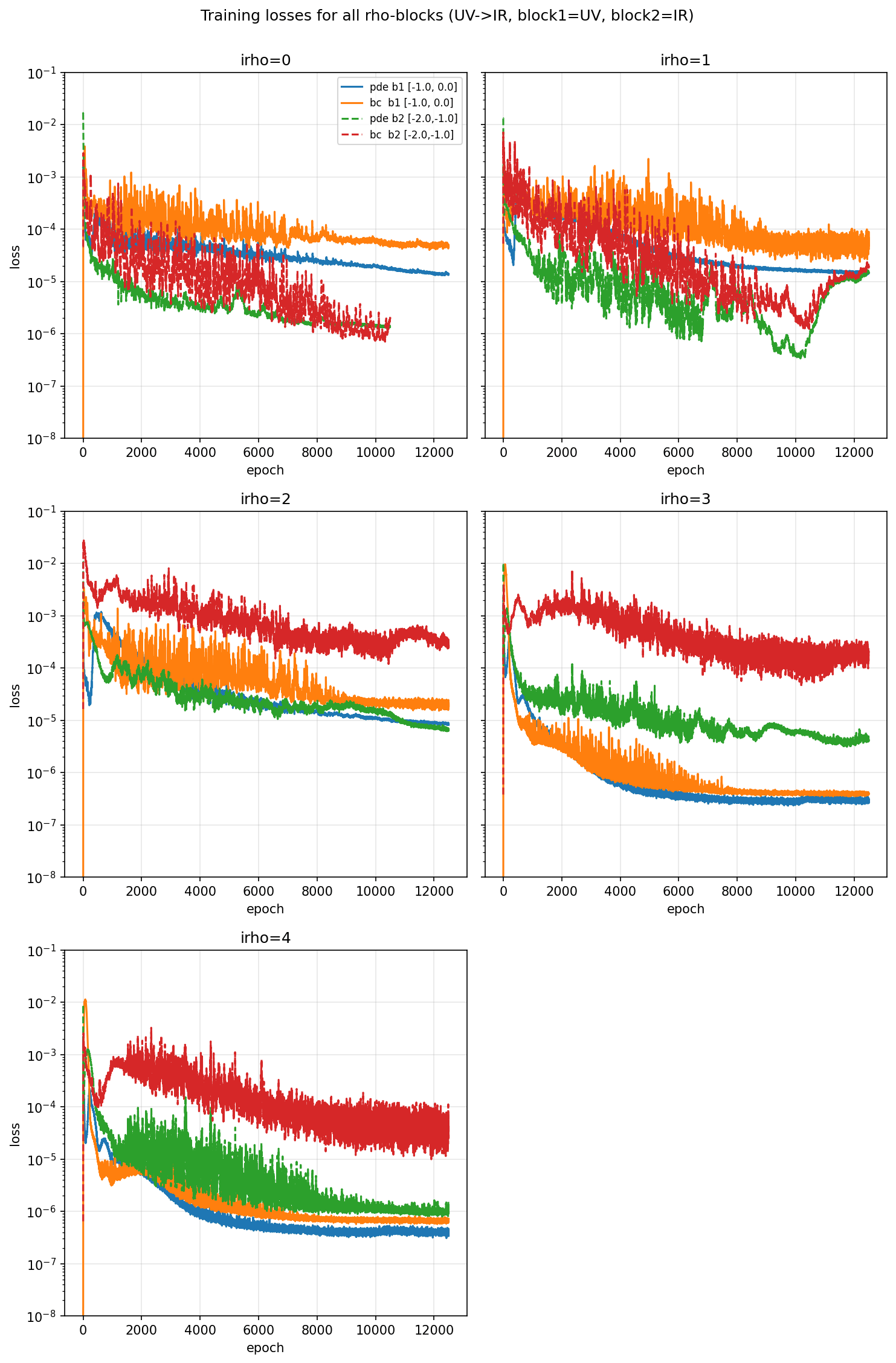}
\caption{Training-loss histories (PDE residual and boundary/interface
condition terms, Eq.~\eqref{eq:loss-total}) for the five $\rho$-direction
field-space blocks of Sec.~\ref{sec:domain-decomp}, at $T=200~\mathrm{GeV}$, on a log scale.
Block~1 covers the UV half of the RG-time range $t\in[-1,0]$ and block~2 the IR
half $t\in[-2,-1]$.}
\label{fig:pinn_training_history}
\end{figure}

Figure~\ref{fig:pinn_training_history} shows a representative example of the
resulting per-block training curves, for the five $\rho$-direction blocks at
$T=200~\mathrm{GeV}$ (the $\sigma$-direction and the other temperature behave
similarly and are omitted here). The PDE and boundary-condition losses of each
block fall by two to four orders of magnitude over the course of training. The
occasional spikes visible for the outer blocks ($i_\rho=2$--$4$) fall at the
boundaries between the short refinement rounds, where the widening of the
$\mathcal{L}_{\rm mag}$ acceptance band shifts the penalty landscape
discontinuously; the learning rate itself is held fixed across these rounds.
After each spike the loss relaxes back down to, or below, its pre-spike level.

\FloatBarrier
\section{Results}
\label{sec:results}

\FloatBarrier
\subsection{A first check: the Higgs-only reduction (PINN)}
\label{sec:results-higgs-only}

We begin, as with the grid-based solver in Sec.~\ref{sec:relax-higgs-only}, with
the simplest version of the problem: the singlet-free reduction of the model --
a single field direction $\rho$, no radial mass matrix and no $\sqrt{\rm disc}$
anywhere. This isolates the role of the soft perturbative-consistency penalty of
Sec.~\ref{sec:loss}, which was motivated in part by the singlet-sector eigenvalue mixing,
from any singlet-specific structure: if the network still depends on the penalty
in this reduction, that dependence cannot be an artifact of the $2\times2$ mass
matrix. We retrained the network for the reduced model at $T=200~\mathrm{GeV}$,
with and without the penalty ($w_{\rm sign}=w_{\rm mag}=0$ in
Eq.~\eqref{eq:loss-total}), keeping the architecture, the multi-block
decomposition, the training schedule and the collocation sampling unchanged.

\begin{figure}[t]
\centering
\includegraphics[width=0.95\linewidth]{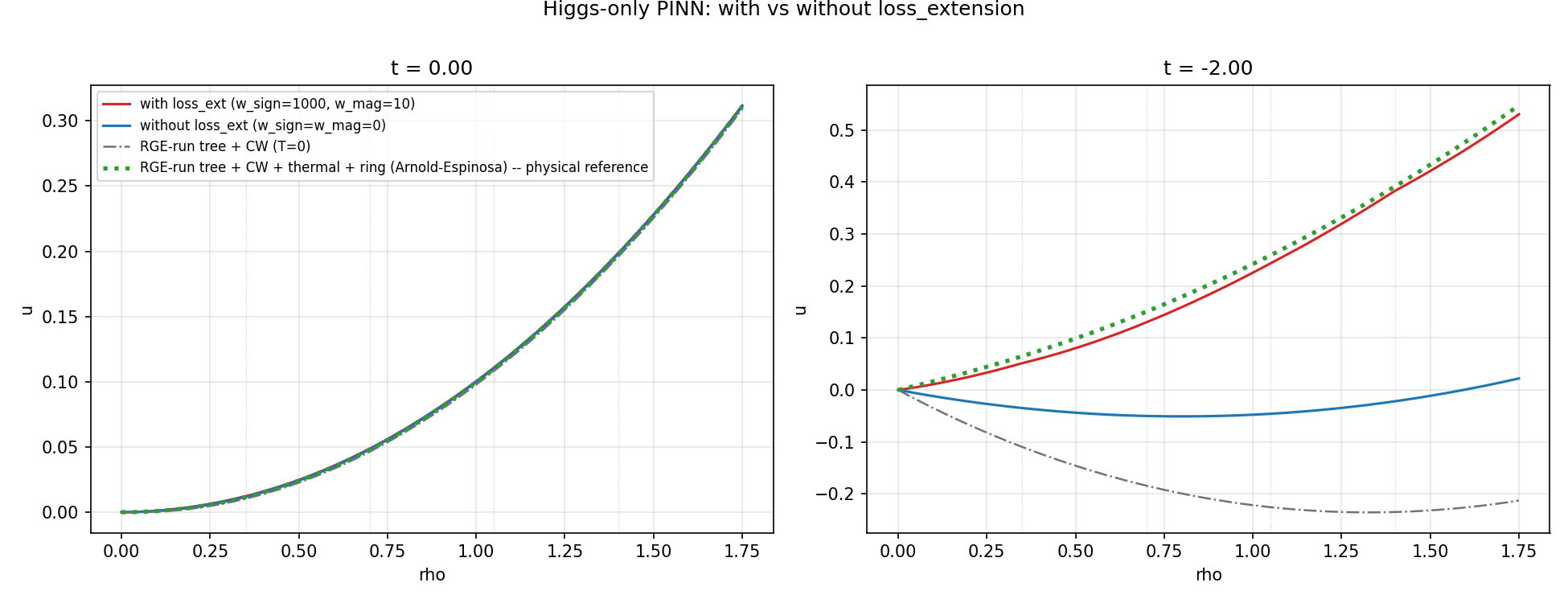}
\caption{First check of the PINN on the Higgs-only reduction at
$T=200~\mathrm{GeV}$: the stitched network prediction (with the gauge-sector ring
correction of Sec.~\ref{sec:ring} added back) along $\rho$ at the UV boundary $t=0$ (left)
and the IR end of the flow $t=-2$ (right), trained with the soft
perturbative-consistency penalty of Sec.~\ref{sec:loss} (red) and without it,
$w_{\rm sign}=w_{\rm mag}=0$ (blue). The penalized network follows the
RGE-improved, Arnold--Espinosa ring-resummed finite-temperature reference (green
dotted) throughout the flow; the unpenalized network drifts to a spurious
nearly-flat solution that tracks neither the finite-temperature nor the
zero-temperature Coleman--Weinberg reference (grey dash-dotted).}
\label{fig:pinn_higgs_only}
\end{figure}

The result is shown in Fig.~\ref{fig:pinn_higgs_only}, at the UV boundary $t=0$
and the IR end of the flow $t=-2$. With the penalty, the
network tracks the ring-resummed finite-temperature reference closely: at $t=-2$
it agrees with the reference to within $\sim3\%$ over the
full field range ($u=0.53$ versus $0.55$ at $\rho=\rho_{\max}=1.75$, and
similarly at every intermediate $\rho$). Without the penalty,
training on the PDE, consistency and boundary/interface losses alone collapses
onto a spurious, nearly flat solution -- $u\simeq0$ across the domain at $t=-2$,
even dipping slightly negative at intermediate $\rho$ -- that matches neither the
finite-temperature reference nor the zero-temperature Coleman--Weinberg curve.
The two solutions are essentially indistinguishable at the UV boundary $t=0$ and
diverge progressively as the flow is integrated toward the IR.

Since this reduced system has none of the singlet mass-matrix structure, the
failure of the unconstrained network to find the physical branch, and the
effectiveness of the loose penalty in restoring it, cannot be attributed to the
eigenvalue-crossing non-analyticity. Both are properties of the nonlinear
loop-plus-thermal flow equation itself and are already present with a single
field: the obstruction is a steep-but-smooth regime of the equation, amplified
toward the IR by the $-4u$ term and the dimensionless rescaling, rather than a
genuine singularity. The full-system slices of Sec.~\ref{sec:results-slices} are
read the same way -- the residual deviations there, along both field directions,
reflect the accuracy a PINN of this size and training budget reaches on an
equation this nonlinear and carry no signature of the eigenvalue-crossing
non-analyticity. The grid-based solver behaves the
same way in this reduction (Sec.~\ref{sec:relax-higgs-only}): without the
coupling prior it produces an untrustworthy solution, and the prior helps but,
unlike here, does not bring it to the finite-temperature answer -- a contrast we
return to in Sec.~\ref{sec:results-limitations}.

\FloatBarrier
\subsection{Slice comparison with the perturbative and relaxation benchmarks}
\label{sec:results-slices}

\begin{figure}[t]
\centering
\begin{subfigure}[t]{0.85\linewidth}
\centering
\includegraphics[width=\linewidth]{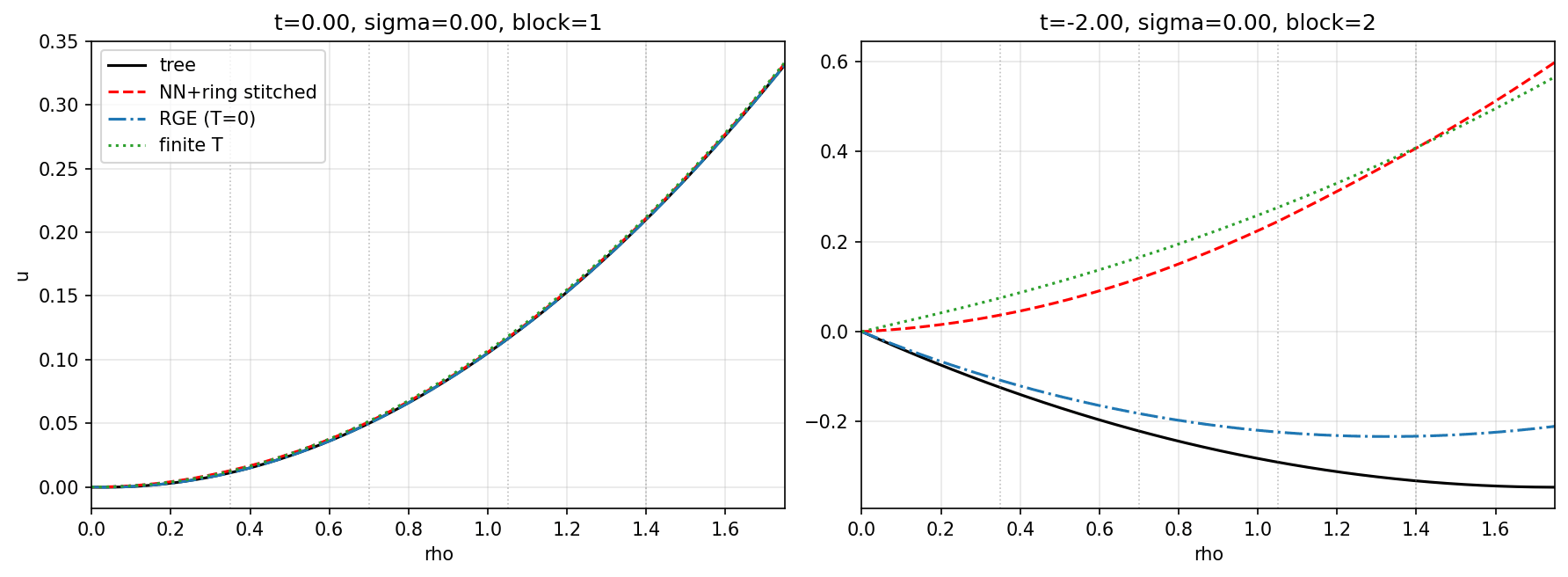}
\caption{$\rho$-direction.}
\label{fig:pinn_rho_T200}
\end{subfigure}

\vspace{1em}

\begin{subfigure}[t]{0.85\linewidth}
\centering
\includegraphics[width=\linewidth]{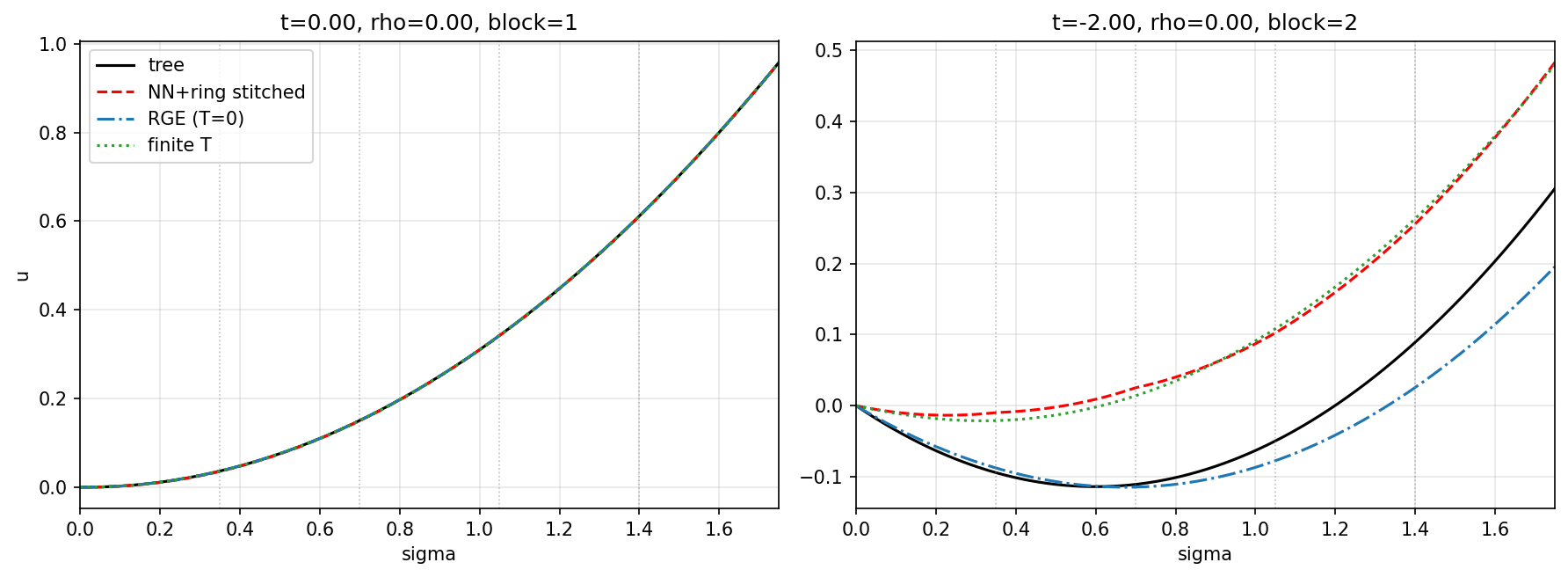}
\caption{$\sigma$-direction.}
\label{fig:pinn_sigma_T200}
\end{subfigure}
\caption{PINN prediction along the $\rho$-direction ($\sigma=0$ fixed, top) and the
$\sigma$-direction ($\rho=0$ fixed, bottom) at $T=200~\mathrm{GeV}$. Each panel shows
the UV boundary $t=0$ (left) and the IR end of the flow $t=-2$ (right): tree-level
potential (black), the stitched network prediction with the gauge-sector ring
correction of Sec.~\ref{sec:ring} added back (red dashed, ``NN+ring stitched''), the
RGE-improved zero-temperature reference (blue dash-dotted), and the same reference
with the finite-temperature piece added (green dotted). Vertical dotted lines mark
the boundaries between the five field-space training blocks of Sec.~\ref{sec:domain-decomp}.}
\label{fig:pinn_slices_T200}
\end{figure}

\begin{figure}[t]
\centering
\begin{subfigure}[t]{0.85\linewidth}
\centering
\includegraphics[width=\linewidth]{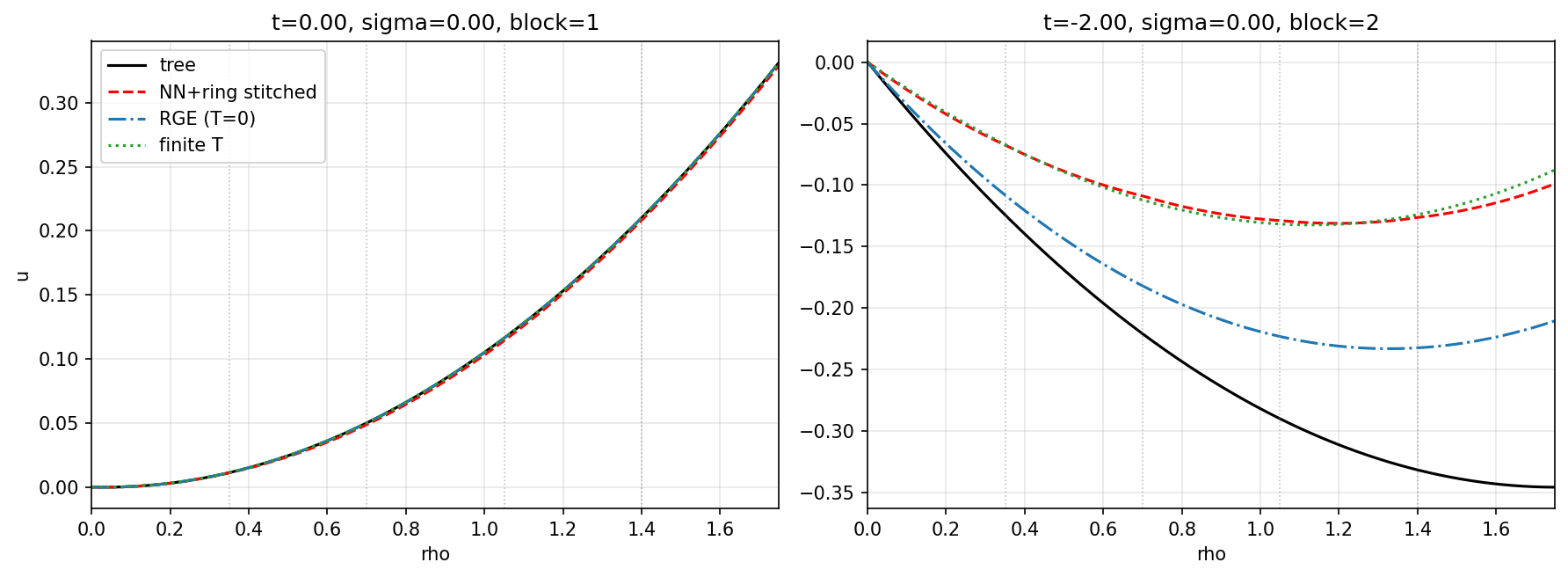}
\caption{$\rho$-direction.}
\label{fig:pinn_rho_T100}
\end{subfigure}

\vspace{1em}

\begin{subfigure}[t]{0.85\linewidth}
\centering
\includegraphics[width=\linewidth]{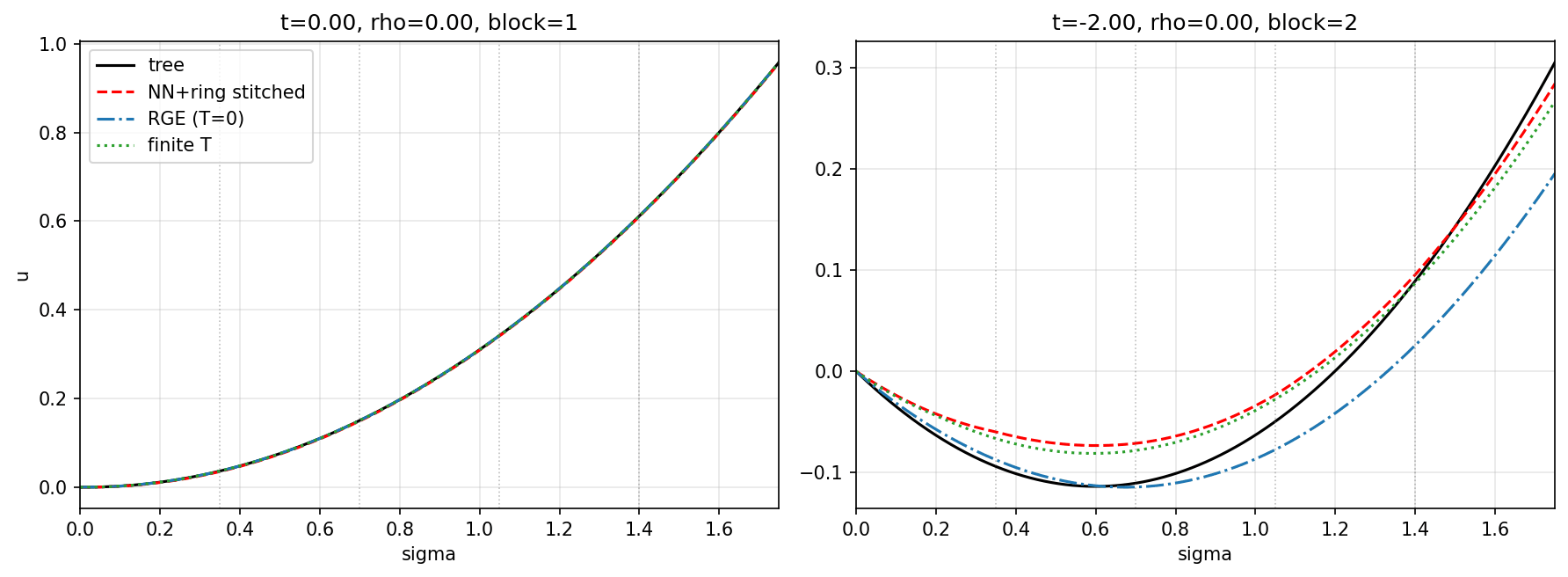}
\caption{$\sigma$-direction.}
\label{fig:pinn_sigma_T100}
\end{subfigure}
\caption{Same as Fig.~\ref{fig:pinn_slices_T200}, at $T=100~\mathrm{GeV}$.}
\label{fig:pinn_slices_T100}
\end{figure}

As a check of the trained network against the same physics used to benchmark the
grid-based solver in Sec.~\ref{sec:relax-benchmark}, Figs.~\ref{fig:pinn_slices_T200} and
\ref{fig:pinn_slices_T100}
show slices of the PINN's potential $u(t,\rho,\sigma)$ along the $\sigma$-direction
($\rho=0$ fixed) and the $\rho$-direction ($\sigma=0$ fixed) -- the true field-space
origin, since the multi-block training of Sec.~\ref{sec:domain-decomp} covers the full domain down to
$\rho=\sigma=0$, unlike the domain-restricted relaxation solver of Sec.~\ref{sec:relax-convergence} -- at the
two RG-time endpoints $t=0$ (the UV boundary, where the network is anchored to the
analytic seed potential of Eq.~\eqref{eq:useed}) and $t=t_{\rm range}=-2$ (the IR end
of the flow), for the same two thermal benchmark points $T=100$ and
$T=200~\mathrm{GeV}$ used in Sec.~\ref{sec:relax-comparison}. Each panel compares four curves: the exact
tree-level potential (black), the network's raw prediction stitched continuously
across the five field-space training blocks of Sec.~\ref{sec:domain-decomp}, with the piecewise
per-block shifts of Sec.~\ref{sec:domain-decomp} fixed by continuity and an overall additive constant
fixed by $u(t,0,0)=0$ (red dashed, ``NN+ring stitched''), a reference built from the
tree-level potential with RG-improved (two-loop running) couplings, augmented only at
$t=t_{\rm range}$ by a static one-loop Coleman--Weinberg correction (blue
dash-dotted, ``RGE $(T=0)$''), and the same RG-improved tree supplemented instead by
the Arnold--Espinosa-resummed finite-temperature one-loop potential (green dotted,
``finite T'') -- the closest analogue, among the reference curves available here, to
the nonperturbative trajectory the network is trying to reproduce. Following the
argument of Sec.~\ref{sec:ring}, only the external gauge sector's ring (Debye-daisy) correction
is added back explicitly on top of the raw network output to form the ``NN+ring''
curve, since the scalar sector's own ring resummation is already generated
internally, nonperturbatively, by the FRG flow itself; the ``finite T'' reference, by
contrast, being built entirely from perturbation theory, must include the scalar
Goldstone/radial ring corrections by hand as well, in addition to the gauge ones.

At $t=0$ all curves coincide with the tree-level potential essentially exactly by
construction, since the UV boundary condition pins the network (and the perturbative
references) to the same seed. The nontrivial comparison is therefore at $t=-2$. At
$T=100~\mathrm{GeV}$ the network reproduces the finite-temperature reference closely
along both field-space axes: the RMS deviation across the slice is below $1\%$ of the
reference's variation along the singlet ($\sigma$) direction and about $3\%$ along the
Higgs ($\rho$) direction (at the field-space maxima, $u=0.28$ against $0.27$ and
$u=-0.10$ against $-0.09$ respectively, the latter pair both small against the tree
value $-0.35$ there). At $T=200~\mathrm{GeV}$ the agreement loosens to an RMS
deviation of $5$--$6\%$ of the slice range in both directions, with the network
running below the reference through the interior of each slice (crossing it near the
outer edge along $\rho$); at the field maxima $u=0.60$ against $0.57$ along $\rho$ and
$u=0.43$ against $0.48$ along $\sigma$. Here the $\sigma$-direction is the singlet
direction (Higgs field held at zero) and the $\rho$-direction the Higgs direction.

The singlet direction is reproduced somewhat more accurately than the Higgs
direction, at both temperatures and by the grid-based relaxation solver of
Sec.~\ref{sec:relax-comparison} as well. We do not read anything structural into
this: the two field directions are inequivalent already at the Lagrangian level --
the Higgs carries the gauge and top-Yukawa loops and a different quartic and mass
parameter from the singlet -- so there is no symmetry between them for the numerics
to respect, and a difference in how accurately each is resolved is unremarkable. It
is in particular not a signature of the ${\rm disc}\to0$ eigenvalue-crossing
non-analyticity of Sec.~\ref{sec:relax-convergence}: the collocation points for these
blocks stay well away from any $1+\bar m^2=0$ threshold, restricting the field range
away from ${\rm disc}=0$ leaves the $\rho$-direction offset essentially unchanged, and
the Higgs-only reduction of Sec.~\ref{sec:results-higgs-only} shows the same residual
difficulty with the singlet, and $\sqrt{\rm disc}$, removed entirely.

The residual deviation from the finite-temperature reference, small at $T=100$ and
growing to a few percent at $T=200$, is best understood simply as the accuracy a PINN
of this size and training budget attains on a flow equation this strongly nonlinear
-- of a piece with the network's documented need for the soft consistency penalty and
hand-tuned loss weights (Secs.~\ref{sec:loss},~\ref{sec:results-limitations}), and
with the stall of the Newton--Krylov iteration on the grid-based side
(Sec.~\ref{sec:relax-convergence}). It does not call for a separate physical or
analytic explanation. We have not carried out a systematic scan in network width or
training budget to confirm this attribution, which we leave to future work.

Across the block interfaces the matched first derivative ($u_\rho$ in the $\rho$
direction, $u_\sigma$ on the singlet slices) agrees between neighboring blocks to
about $2\%$ of the local slope in the median and to under $\sim\!10\%$ at the
ninetieth percentile -- a few $\times 10^{-3}$ in absolute terms, comparable to the
pointwise deviations from the reference quoted above -- at both benchmark
temperatures, with larger relative values confined to the immediate neighborhood of
the origin at the deepest infrared times, where the slope itself passes through zero.
The interface conditions thus enforce $C^1$ continuity to within the overall solution
accuracy rather than exactly.

The derivative heads themselves, tied to the true derivatives of the scalar output
only softly through $\mathcal{L}_{\rm consist}$ (Sec.~\ref{sec:loss}), stay close to
them in the converged networks: sampled over the collocation domain, the RMS relative
mismatch $\langle(\partial_\rho u-u_\rho)^2\rangle^{1/2}/\langle u_\rho^2\rangle^{1/2}$
between each head and one automatic differentiation of the next-lower head is below
$0.4\%$ for every first- and second-derivative head, and under $0.7\%$ in the worst
individual block, at both benchmark temperatures -- several times tighter than the median
interface $C^1$ mismatch and well below the deviation from the resummed-perturbation
reference.

\FloatBarrier
\subsection{Toward the two-dimensional potential}
\label{sec:results-2d}

The results above are all restricted to one-dimensional slices in field space --
the axes $\sigma=0$ and $\rho=0$ of the full system
(Sec.~\ref{sec:results-slices}), and the singlet-free reduction of
Sec.~\ref{sec:results-higgs-only} -- which is sufficient to fix the location and depth
of each vacuum separately but does not resolve the potential barrier
between them -- the quantity that ultimately controls whether the
electroweak step of a two-step transition (Sec.~\ref{sec:vacuum}) is strongly
first-order. Assessing that barrier requires the full two-dimensional
potential $u(t,\rho,\sigma)$, and as a first step in this direction we
have extended the multi-block architecture of Sec.~\ref{sec:domain-decomp} from the two
one-dimensional slices to a genuine $5\times5$ grid of blocks in
$(\rho,\sigma)$, stitched together with the same continuity matching used
in one dimension. Training of this 2D network is at present only
preliminary and has not been tuned to the same accuracy as the
one-dimensional results reported above; the example shown here should
therefore be read as a demonstration of the framework rather than as a
converged physics result. The harder convergence of the two-dimensional run
is therefore not a qualitative jump to a more complex equation -- it is the
same flow equation and architecture already exercised by the one-dimensional
runs (Sec.~\ref{sec:domain-decomp}), now tiled over the full domain in both
directions ($5\times5$ instead of $5\times1$ blocks) rather than a narrow slab
in the secondary field, and consequently exposed to the genuinely mixed,
far-from-axis region that the one-dimensional runs never had to resolve.

\begin{figure}[t]
\centering
\includegraphics[width=0.75\linewidth]{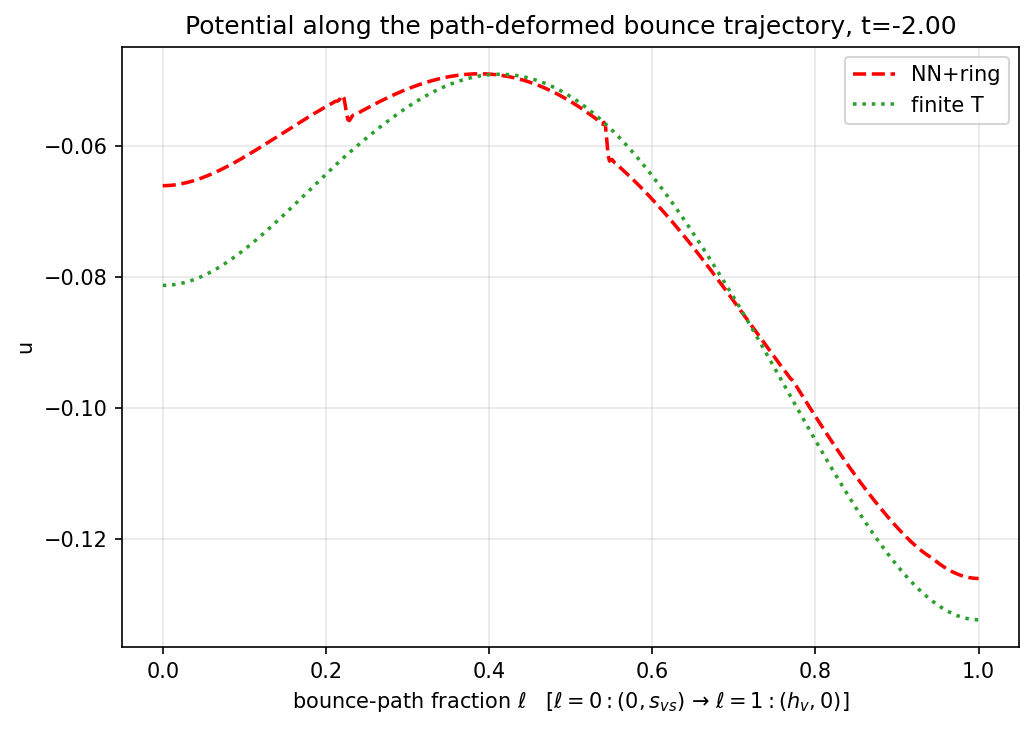}
\caption{Preliminary two-dimensional PINN prediction (with the gauge-sector
ring correction of Sec.~\ref{sec:ring} added back, ``NN+ring'') for the
potential along the path-deformed $O(3)$ bounce trajectory connecting the
singlet-only vacuum $(h,s)=(0,v_S)$ and the electroweak vacuum
$(h,s)=(v,0)$, at $t=-2$ and $T=100~\mathrm{GeV}$, parametrized by
normalized arc length $\ell$, with $\ell=0$ at the singlet-only vacuum and
$\ell=1$ at the electroweak vacuum. The ``NN+ring'' and ``finite~$T$''
curves are each drawn along the deformed bounce trajectory in their own
potential, connecting that potential's own pair of axis vacua (both
located from the corresponding 2D prediction; these are the same
trajectories entering the $S_3$ computation in the text). The 2D training
underlying this figure
is not yet converged to the same accuracy as the one-dimensional results
above; see the discussion in the text.}
\label{fig:pinn_vacuum_line}
\end{figure}

Figure~\ref{fig:pinn_vacuum_line} shows the potential along the
path-deformed $O(3)$ bounce trajectory connecting the two axis vacua
$(h,s)=(0,v_S)$ and $(h,s)=(v,0)$ at $t=-2$ and $T=100~\mathrm{GeV}$ --
that is, the same trajectory obtained by minimizing the bounce action
below (Sec.~\ref{sec:tc-tn}, via the deformation of
Ref.~\cite{Wainwright:2011kj}), rather than a straight line imposed by
hand. The ``NN+ring'' and ``finite~$T$'' curves are each drawn along
their own bounce trajectory in the respective potential, closed onto that
potential's own pair of axis vacua (both located from the corresponding
2D prediction); the potential accordingly approaches each endpoint
quadratically, with vanishing slope, although the flat region is narrow
on the scale of the plot because the wells are steep in the physical
amplitudes $(h,s)$. The resulting curve already
reproduces the qualitative shape expected for a two-step history -- a
single potential barrier separating the singlet-broken and electroweak
minima -- although the height and asymmetry of the barrier still differ
visibly from the finite-temperature reference (the network's barrier,
$u_{\max}-u_{\rm false}\approx0.017$, is about half the reference's
$\approx0.032$), consistent with the
preliminary state of the training noted above.

With a barrier resolved, the framework already delivers a
nucleation-scale observable end to end. Taking the flowed two-dimensional
potential of Fig.~\ref{fig:pinn_vacuum_line} as input, we solve the $O(3)$
bounce equation of Sec.~\ref{sec:tc-tn} along the path-deformed
trajectory~\cite{Wainwright:2011kj} connecting the two vacua, with the
potential and its gradient supplied analytically by the trained network
and the kinetic terms taken canonically normalized ($Z_H=Z_S=1$) in the
bounce functional, i.e.\ dropping the wave-function-renormalization running
in this final step. Dropping the $\eta$
running here is well justified: $\eta_H\simeq\tfrac{2}{16\pi^2}(3y_t^2-\tfrac94 g_2^2-\tfrac34 g_1^2)\approx0.02$
so $Z_H$ varies by only a few percent over $|t|\le2$, $\eta_S\sim10^{-4}$ is
negligible, and the resulting shift in $S_3/T$ is far below the
barrier-convergence uncertainty that dominates the quoted number. This gives
$S_3(T)/T \approx \SthreeDeformed$ at $T=100~\mathrm{GeV}$, against
$S_3(T)/T \approx \SthreeRef$ obtained from the perturbative
Arnold--Espinosa finite-temperature reference potential evaluated the same
way -- a factor of two below the reference, consistent with the
still-visible barrier-height difference between the ``NN+ring'' and
``finite $T$'' curves of Fig.~\ref{fig:pinn_vacuum_line}. Both values sit well
above the nucleation criterion $S_3(T_n)/T_n\simeq140$ of
Sec.~\ref{sec:tc-tn}, so this benchmark places $T_n$ below
$100~\mathrm{GeV}$; the un-converged barrier and the absence of a temperature
scan preclude a quantitative determination of $T_n$ itself. A straight-line
single-field instanton bound on $S_3$, attempted as a cross-check, did not
converge numerically for this network snapshot and is not quoted. The remaining
gap is dominated by the still-incomplete convergence of the 2D barrier
rather than by the bounce computation; with the present training we
cannot separate this convergence error from any genuine nonperturbative
reduction of the barrier along the flow. We therefore
quote the number as an order-of-magnitude, end-to-end demonstration rather than a
converged transition strength.
Bringing the 2D network to the same level of convergence as the 1D
slices, propagating the residual $k_{\rm IR}$-truncation systematic, and
turning this into a determination of $T_n$ and $\xi_c$, is left for
future work.

\FloatBarrier
\subsection{Practical role and limitations of the soft consistency penalty}
\label{sec:results-limitations}

As anticipated in Sec.~\ref{sec:loss}, the soft perturbative-consistency penalty
($\mathcal{L}_{\rm sign}$ and $\mathcal{L}_{\rm mag}$) turned out to be necessary
rather than optional: without it, training on the PDE residual, consistency, and
boundary/interface losses alone did not reliably converge to a physically
sensible solution, and reaching the results above also required hand-tuning the
relative weights of $\mathcal{L}_{\rm sign}$ and $\mathcal{L}_{\rm mag}$ across
the constrained combinations (Appendix~\ref{app:impl}), with weights that
differed between the two benchmark temperatures. We report this openly as a practical
limitation of the present setup.

What the penalty actually enforces is weak. It fixes only the sign and the order of
magnitude of a few local low-order coefficients of the potential: the penalized mass
combination -- the singlet $a_S$, and at $T=100\gev$ also the Higgs $a_H$ -- has its
shift away from tree level held between roughly a tenth and three times the corresponding
analytic (RGE-running plus thermal) shift, and the quartic curvatures are held within
about a factor of two of their tree-level (inner region) or RGE-running (outer region)
value. Nothing is imposed on the height or position of the barrier, on the location or
depth of the infrared minima, or on the potential away from the two RG-time slices
$t=-1,-2$ on which the penalty acts. In the ideal case -- a weighting on which the
converged solution does not depend -- the penalty would then act purely as a branch
selector: of the several nearby near-stationary points that the reduced loss admits for
this nonlinear flow (Sec.~\ref{sec:loss}), it would pick out the one consistent with
perturbation theory in sign and magnitude, without dictating the result, so that the
reported potential remains a solution of the flow equation, simply the physically
sensible one. The limitation documented below is that this ideal is not quite reached in
practice: the converged infrared potential keeps a residual, quantitatively relevant
dependence on the hand-tuned weights and on the analytic thermal target.

It is worth being precise about where this difficulty actually sits. Both
methods handle the tree-level problem without any such guidance: the
tree-only relaxation system is linear and solved essentially exactly by a
single LU factorization (Sec.~\ref{sec:relax-method}), and the PINN likewise reproduces the
tree-level potential to high accuracy at the UV boundary $t=0$ by
construction (see above). The breakdown in both cases is specifically tied
to switching on the loop contributions -- the $E_{\rm ext}$ term and the
$\coth$ threshold functions of Eq.~\eqref{eq:flow-dimless}, including their
finite-temperature part -- which is what makes the flow equation genuinely
nonlinear and is exactly where the Newton--Krylov iteration stalls
(Sec.~\ref{sec:relax-convergence}) and where the PINN requires the soft consistency penalty to stay
on a physically sensible branch. The comparison between the two methods
should therefore be read as a comparison of how each one copes once loop
(including thermal) effects are included, not of the tree-level problem,
which both solve comfortably on their own.

At the same time, this is a meaningful contrast with, rather than a
repetition of, the outcome of Sec.~\ref{sec:relax-benchmark}: there, an essentially analogous
coupling-prior forcing term was added directly to the Newton--Krylov
residual (Sec.~\ref{sec:coupling-prior}), and it did \emph{not} rescue the
solver -- turning it on drove the final residual norm up rather than down in
three of the four solver/stencil pairs (Table~\ref{tab:relax}), and in the one
pair where it helped only by regressing the worst-converged run toward the same
plateau as the rest, so the grid-based
method could not be brought to a physically sensible solution either with
or without this guidance. For the PINN, by contrast, adding the same kind
of loose perturbative guidance, once its relative weights were tuned, did
yield the physically sensible, closely tree/finite-temperature-tracking
solutions reported above. So while neither method solves the flow equation
from the PDE residual and boundary/interface conditions alone, only the
PINN was able to make productive use of the additional perturbative
guidance once supplied; the rigidity of the fixed finite-difference
stencil, rather than the idea of a coupling prior itself, appears to be
what prevented the relaxation solver from benefiting in the same way. We
regard removing the PINN's remaining dependence on hand-tuned weights,
whether through a better-conditioned network architecture, a more
informative choice of collocation points, or a principled rather than
hand-tuned weighting of the consistency terms, as an important open problem
rather than a solved part of the present framework.

The dependence runs deeper than the choice of weights. In the
finite-temperature runs the mass-like combinations $a_H,a_S$ are compared
against an analytic thermal target obtained by differentiating the
one-loop ring-resummed thermal potential (Sec.~\ref{sec:loss}), and we find
the converged infrared potential to retain a visible sensitivity both to
that target and to which combinations receive a penalty at all -- beyond
the sensitivity to the weights and to the on/off switch already discussed.
The agreement of the mass sector with the perturbative finite-temperature
reference is therefore in part imposed through this target rather than a
pure output of the flow. The genuinely unconstrained, nonperturbative
content of the solution sits in the quartic/shape sector, which carries
only the loose cap $|\Delta\lambda_i^{\rm NN}/\lambda_{i,\rm tree}|\lesssim1$
-- and this is precisely the sector that is not yet robust against the
above choices. Making the flow equation together with the genuine UV and
thermal boundary data fix the mass sector on their own, without an analytic
thermal target, is in our view the central open problem, ahead of raw
network capacity.

\FloatBarrier
\section{Conclusions and discussion}
\label{sec:conclusions}

We have applied physics-informed neural networks to the Wetterich flow
equation of a $Z_2$-symmetric singlet extension of the Standard Model,
including SM gauge/Yukawa loops and finite-temperature effects, and used the
resulting effective potential to demonstrate an end-to-end electroweak-phase-transition
pipeline. A mesh-free, multi-block PINN architecture reproduces the
tree-level potential essentially exactly at the UV boundary by construction,
and at the IR end of the flow tracks the perturbative (RGE-improved,
finite-temperature) reference to within a few percent along both field-space
axes -- most closely at $T=100~\mathrm{GeV}$, with the deviation growing to
$5$--$6\%$ at $T=200~\mathrm{GeV}$, at the level of accuracy expected of a PINN
on an equation this nonlinear (Sec.~\ref{sec:results}). As a
preliminary step toward the two-dimensional potential needed to assess the
barrier between the singlet-broken and electroweak minima, we also trained a
$5\times5$-block extension of the network and used it to locate both vacua
directly and evaluate the potential along the line connecting them; while
not yet converged to the same accuracy as the one-dimensional results, it
already reproduces the qualitative barrier structure expected of a two-step
transition.

To assess whether this outcome required the flexibility of a neural network
in particular, or would have followed from discretizing the same flow
equation by more conventional means, we benchmarked the PINN against a
grid-based Newton--Krylov relaxation solver applied to the identical
equation (Sec.~\ref{sec:relax-benchmark}). The comparison is informative precisely because it is
asymmetric: the relaxation solver does not converge to a physically sensible
solution once loop and thermal effects are switched on, and adding a
coupling-prior forcing term analogous to the PINN's soft
perturbative-consistency penalty does not rescue it -- in most of the runs it
drives the residual up rather than down (Table~\ref{tab:relax}). The PINN, by
contrast, does reach physically sensible solutions once the analogous penalty is
included, though only after hand-tuning its relative weights. The rigidity of the
fixed finite-difference stencil, rather than the coupling-prior idea itself,
appears to be what prevents the grid method from benefiting from the same
guidance (Sec.~\ref{sec:results-limitations}).

Repeating both calculations for a singlet-free reduction of the model --
Higgs plus the external gauge and top sector only, with no $2\times2$ radial
mass matrix and hence none of the near-degenerate eigenvalue mixing of the
full system -- shows that this contrast does not
originate in the singlet sector
(Secs.~\ref{sec:relax-higgs-only},~\ref{sec:results-higgs-only}). In the
reduction the grid solver behaves better -- it can even reach tolerance -- but
still converges to a spurious rather than the physical solution, while the
unconstrained PINN still collapses to a nearly-flat solution and still needs the
soft penalty to track the finite-temperature reference. The dependence on perturbative guidance, and
the difference in how the two methods make use of it, are therefore
properties of the nonlinear loop-plus-thermal flow equation itself rather
than artifacts of the Higgs--singlet mass mixing.

As a first end-to-end test of the pipeline we also solved the $O(3)$
bounce equation on the preliminary 2D potential, obtaining
$S_3(T)/T\approx\SthreeDeformed$ at $T=100~\mathrm{GeV}$ against
$\approx\SthreeRef$ for the perturbative finite-temperature reference; the
gap is dominated by the un-converged 2D barrier and the number is quoted
only as a demonstration (Sec.~\ref{sec:results-2d}).

A quantitative comparison against an established nonperturbative result is
not yet within reach for this model, on two counts. First, there is no
truncation-free benchmark: the most complete existing nonperturbative
treatment, a three-dimensional-EFT lattice study of the same $Z_2$ singlet
model~\cite{Niemi:2024xsm}, itself relies on a perturbative dimensional
reduction to build the effective theory it then simulates. Second, and more
fundamentally, our infrared potential is not yet independent of the
perturbative guidance used in training: as discussed in
Sec.~\ref{sec:results-limitations}, the mass sector is tied to an analytic
thermal target, so comparing that sector against a perturbative or lattice
benchmark would be partly circular, while the quartic/shape sector, though
unconstrained, is not yet robust enough to compare quantitatively.
Decoupling the flow result from the perturbative targets and then
benchmarking it against Ref.~\cite{Niemi:2024xsm} is left for future work.

Several directions follow naturally from this work. On the numerical side,
bringing the two-dimensional PINN to the same level of convergence as the
one-dimensional results, and using its analytic barrier and bounce action
along general paths in field space, would turn this demonstration into a
fully nonperturbative determination of the nucleation temperature and
transition strength. Removing the PINN's remaining dependence on hand-tuned
consistency-loss weights -- whether through a better-conditioned
architecture, a more informative choice of collocation points, or a
principled rather than hand-tuned weighting scheme -- remains an open
problem. On the physics side, extending the present $Z_2$-symmetric,
CP-conserving truncation to include the CP-violating interactions needed for
a complete treatment of electroweak baryogenesis is a natural next step, as
is applying the same mesh-free FRG approach to other multi-field BSM
extensions where a fully nonperturbative effective potential is needed.

\section*{Code and data availability}
	The source code for the physics-informed neural network training, the
	grid-based Newton--Krylov relaxation solver, and the scripts that generate
	the figures in this paper are openly available at
	\url{https://github.com/nyokozaki/frg_pinns_higgs_singlet_pub}.
	The trained network weights, the loss-weight configurations, and the
	benchmark input parameters used for the results reported here are attached to
	the tagged \texttt{v1.0} release of that repository.

\section*{Acknowledgments}
	This work is supported by a startup grant from Zhejiang University.

\FloatBarrier
\appendix

\FloatBarrier
\section{Implementation details}
\label{app:impl}

Table~\ref{tab:impl} collects the network, optimizer, and loss-function
hyperparameters used for the finite-temperature runs of Sec.~\ref{sec:results}.
Entries that differ between the two benchmark temperatures are written as
$T{=}100\,|\,T{=}200$. The complete per-block configuration files for every run
are included in the code release.

\begin{table}[h]
\centering
\small
\begin{tabular}{@{}l p{0.5\linewidth}@{}}
\hline\hline
Quantity & Value \\
\hline
\multicolumn{2}{l}{\textit{Network} (\texttt{BaseNet}, one per domain block)} \\
\hline
stem MLP (layers $\times$ width)        & $3\times160$ \\
$u$-branch MLP                          & $1\times160$ \\
$\rho$-, $\sigma$-branch MLP            & $2\times160$ \\
mixed-branch MLP                        & $2\times320$ \\
activation / output heads               & SiLU / linear, zero-initialized \\
parameters per \texttt{BaseNet}         & $\approx3.9\times10^{5}$ \\
blocks (1D / 2D)                        & $5\times2$ / $5\times5\times2$ \\
field-block width $\Delta\rho$ / overlap & $0.35$ / $0.02$ \\
\hline
\multicolumn{2}{l}{\textit{Baseline decomposition} $u=\tilde u_{\rm tree}+\tilde u_{\rm th}+N$ (finite $T$)} \\
\hline
$\tilde u_{\rm tree}$ rescalings of the 5 tree terms & $(1,1,0.8,0.8,0.8)$ \\
$\tilde u_{\rm th}$ coefficients $(c_{T,\rho},\,c_{T,\sigma})$ & $(0.2,\,0.1)$ \\
\hline
\multicolumn{2}{l}{\textit{Optimizer and schedule} (per block)} \\
\hline
optimizer                              & SOAP~\cite{SOAP} \\
base learning rate $\to\eta_{\min}$     & $3\times10^{-4}\to10^{-5}$, cosine anneal \\
iterations per block                    & $10^{4}$ \\
refinement rounds                       & up to $25\times100$ iter, frozen lr, stop at loss ${<}10^{-9}$ \\
collocation / iter (PDE/BC/interface/penalty) & $7168$ / $4096$ / $2048$ / $1024$ \\
\hline
\multicolumn{2}{l}{\textit{Loss weights}, Eq.~\eqref{eq:loss-total}} \\
\hline
$w_{\rm pde}$, $w_{\rm consist}$, $w_{\rm bc}$ & $1$, $1$, $1$ \\
$w_{s1}=w_{s2}=w_{rs}$ (in $\mathcal{L}_{\rm consist}$) & $5$ \\
second-derivative term in $\mathcal{L}_{\rm bc}$ & $0$ \\
$w_{\rm overlap}$ (1D / 2D)             & $0.05$ / $0.01$ \\
$w_{\rm sign}$, $w_{\rm mag}$           & $10^{3}$, $10$ \\
hinge $\beta$, $(c_-,c_+)$              & $100$, $(0.1,3.0)$ \\
\hline
\multicolumn{2}{l}{\textit{Sign/magnitude penalty content} (finite $T$; $T{=}100\,|\,T{=}200$)} \\
\hline
inner/outer field-space split $(\rho_{\rm cw},\sigma_{\rm cw})$ & $(1.2,0.8)\,|\,(3.0,0.5)$ \\
\emph{inner:} sign hinge on            & $a_S$ (rel.\ weight $1$) \\
\emph{inner:} magnitude band on        & $a_S$ (rel.\ weight $2$); $a_H$ (rel.\ weight $0.1\,|\,0$) \\
\emph{inner:} quartic flat cap $|\Delta\lambda_i/\lambda_{i,\rm tree}|{<}0.99$ & $\lambda_H$: $0.5\,|\,0.1$; $\lambda_S$: $1.0$; $\lambda_{HS}$: $0.01$ \\
\emph{outer:} quartic RGE-target band  & $\lambda_H$: $0.5\,|\,0$; $\lambda_S$: $1.0\,|\,1.0$ \\
\hline\hline
\end{tabular}
\caption{Network, optimizer, and loss hyperparameters for the finite-temperature
PINN runs of Sec.~\ref{sec:results}. ``rel.\ weight'' denotes the weight of a
combination inside the average defining $\mathcal{L}_{\rm sign}$ or
$\mathcal{L}_{\rm mag}$, before multiplication by the outer $w_{\rm sign}$,
$w_{\rm mag}$. Combinations not listed in the last block carry no penalty at
finite temperature.}
\label{tab:impl}
\end{table}

\FloatBarrier
\section{Two-loop renormalization group equations}
\label{app:rge}

The two-loop RGEs used to run the couplings between $k_{\rm IR}$ and $k_{\rm UV}$
(Sec.~\ref{sec:benchmark-point}) and to build the perturbative targets of the soft
consistency penalty (Sec.~\ref{sec:loss}) were generated with
PyR@TE3~\cite{Sartore:2020gyh} for the $Z_2$-symmetric model of
Eq.~\eqref{eq:u-uv}. Their one-loop parts agree with those quoted in the
singlet-extension literature~\cite{Gonderinger:2009jp}, and the two-loop
structure follows the general results of
Refs.~\cite{Machacek:1983tz,Machacek:1983fi,Machacek:1984zw}. We use the
convention
\begin{equation}
\beta(X) \equiv \mu\,\frac{dX}{d\mu}
= \frac{1}{(4\pi)^{2}}\,\beta^{(1)}(X) + \frac{1}{(4\pi)^{4}}\,\beta^{(2)}(X) ,
\end{equation}
keep only the top Yukawa coupling $y_t$ (so $Y_u=\mathrm{diag}(0,0,y_t)$), and
write $\mu_H^2,\mu_S^2$ for the Higgs and singlet mass parameters and
$\lambda_H,\lambda_S,\lambda_{HS}$ for the quartics of Eq.~\eqref{eq:u-uv}.
\allowdisplaybreaks

\paragraph{Gauge couplings.}

\begin{align*}
\begin{autobreak}
\beta^{(1)}(g_1) =\frac{41}{6} g_1^{3}
\end{autobreak}
\end{align*}
\begin{align*}
\begin{autobreak}
\beta^{(2)}(g_1) =

+ \frac{199}{18} g_1^{5}

+ \frac{9}{2} g_1^{3} g_2^{2}

+ \frac{44}{3} g_1^{3} g_3^{2}

-  \frac{17}{6} g_1^{3} y_t^{2}
\end{autobreak}
\end{align*}
\begin{align*}
\begin{autobreak}
\beta^{(1)}(g_2) =- \frac{19}{6} g_2^{3}
\end{autobreak}
\end{align*}
\begin{align*}
\begin{autobreak}
\beta^{(2)}(g_2) =

+ \frac{3}{2} g_1^{2} g_2^{3}

+ \frac{35}{6} g_2^{5}

+ 12 g_2^{3} g_3^{2}

-  \frac{3}{2} g_2^{3} y_t^{2}
\end{autobreak}
\end{align*}
\begin{align*}
\begin{autobreak}
\beta^{(1)}(g_3) =-7 g_3^{3}
\end{autobreak}
\end{align*}
\begin{align*}
\begin{autobreak}
\beta^{(2)}(g_3) =

+ \frac{11}{6} g_1^{2} g_3^{3}

+ \frac{9}{2} g_2^{2} g_3^{3}

- 26 g_3^{5}

- 2 g_3^{3} y_t^{2}
\end{autobreak}
\end{align*}

\paragraph{Top Yukawa coupling.}

\begin{align*}
\begin{autobreak}
\beta^{(1)}(y_t) =

+ \frac{9}{2} y_t^{3}

-  \frac{17}{12} g_1^{2} y_t

-  \frac{9}{4} g_2^{2} y_t

- 8 g_3^{2} y_t
\end{autobreak}
\end{align*}
\begin{align*}
\begin{autobreak}
\beta^{(2)}(y_t) =

- 12 y_t^{5}

- 12 \lambda_H y_t^{3}

+ 6 \lambda_H^{2} y_t

+ \frac{1}{4} \lambda_{HS}^{2} y_t

+ \frac{131}{16} g_1^{2} y_t^{3}

+ \frac{225}{16} g_2^{2} y_t^{3}

+ 36 g_3^{2} y_t^{3}

+ \frac{1187}{216} g_1^{4} y_t

-  \frac{3}{4} g_1^{2} g_2^{2} y_t

+ \frac{19}{9} g_1^{2} g_3^{2} y_t

-  \frac{23}{4} g_2^{4} y_t

+ 9 g_2^{2} g_3^{2} y_t

- 108 g_3^{4} y_t
\end{autobreak}
\end{align*}

\paragraph{Scalar quartic couplings.}

\begin{align*}
\begin{autobreak}
\beta^{(1)}(\lambda_H) =

+ 24 \lambda_H^{2}

+ \frac{1}{2} \lambda_{HS}^{2}

- 3 g_1^{2} \lambda_H

- 9 g_2^{2} \lambda_H

+ \frac{3}{8} g_1^{4}

+ \frac{3}{4} g_1^{2} g_2^{2}

+ \frac{9}{8} g_2^{4}

+ 12 \lambda_H y_t^{2}

- 6 y_t^{4}
\end{autobreak}
\end{align*}
\begin{align*}
\begin{autobreak}
\beta^{(2)}(\lambda_H) =

- 312 \lambda_H^{3}

- 5 \lambda_{HS}^{2} \lambda_H

- 2 \lambda_{HS}^{3}

+ 36 g_1^{2} \lambda_H^{2}

+ 108 g_2^{2} \lambda_H^{2}

+ \frac{629}{24} g_1^{4} \lambda_H

+ \frac{39}{4} g_1^{2} g_2^{2} \lambda_H

-  \frac{73}{8} g_2^{4} \lambda_H

-  \frac{379}{48} g_1^{6}

-  \frac{559}{48} g_1^{4} g_2^{2}

-  \frac{289}{48} g_1^{2} g_2^{4}

+ \frac{305}{16} g_2^{6}

- 144 \lambda_H^{2} y_t^{2}

+ \frac{85}{6} g_1^{2} \lambda_H y_t^{2}

+ \frac{45}{2} g_2^{2} \lambda_H y_t^{2}

+ 80 g_3^{2} \lambda_H y_t^{2}

-  \frac{19}{4} g_1^{4} y_t^{2}

+ \frac{21}{2} g_1^{2} g_2^{2} y_t^{2}

-  \frac{9}{4} g_2^{4} y_t^{2}

- 3 \lambda_H y_t^{4}

-  \frac{8}{3} g_1^{2} y_t^{4}

- 32 g_3^{2} y_t^{4}

+ 30 y_t^{6}
\end{autobreak}
\end{align*}
\begin{align*}
\begin{autobreak}
\beta^{(1)}(\lambda_S) =

+ 18 \lambda_S^{2}

+ 2 \lambda_{HS}^{2}
\end{autobreak}
\end{align*}
\begin{align*}
\begin{autobreak}
\beta^{(2)}(\lambda_S) =

- 204 \lambda_S^{3}

- 20 \lambda_{HS}^{2} \lambda_S

- 8 \lambda_{HS}^{3}

+ 4 g_1^{2} \lambda_{HS}^{2}

+ 12 g_2^{2} \lambda_{HS}^{2}

- 12 \lambda_{HS}^{2} y_t^{2}
\end{autobreak}
\end{align*}
\begin{align*}
\begin{autobreak}
\beta^{(1)}(\lambda_{HS}) =

+ 12 \lambda_{HS} \lambda_H

+ 6 \lambda_{HS} \lambda_S

+ 4 \lambda_{HS}^{2}

-  \frac{3}{2} g_1^{2} \lambda_{HS}

-  \frac{9}{2} g_2^{2} \lambda_{HS}

+ 6 \lambda_{HS} y_t^{2}
\end{autobreak}
\end{align*}
\begin{align*}
\begin{autobreak}
\beta^{(2)}(\lambda_{HS}) =

- 72 \lambda_{HS}^{2} \lambda_H

- 36 \lambda_{HS}^{2} \lambda_S

- 60 \lambda_{HS} \lambda_H^{2}

- 30 \lambda_{HS} \lambda_S^{2}

-  \frac{21}{2} \lambda_{HS}^{3}

+ 24 g_1^{2} \lambda_{HS} \lambda_H

+ 72 g_2^{2} \lambda_{HS} \lambda_H

+ g_1^{2} \lambda_{HS}^{2}

+ 3 g_2^{2} \lambda_{HS}^{2}

+ \frac{557}{48} g_1^{4} \lambda_{HS}

+ \frac{15}{8} g_1^{2} g_2^{2} \lambda_{HS}

-  \frac{145}{16} g_2^{4} \lambda_{HS}

- 72 \lambda_{HS} \lambda_H y_t^{2}

- 12 \lambda_{HS}^{2} y_t^{2}

+ \frac{85}{12} g_1^{2} \lambda_{HS} y_t^{2}

+ \frac{45}{4} g_2^{2} \lambda_{HS} y_t^{2}

+ 40 g_3^{2} \lambda_{HS} y_t^{2}

-  \frac{27}{2} \lambda_{HS} y_t^{4}
\end{autobreak}
\end{align*}

\paragraph{Scalar mass parameters.}

\begin{align*}
\begin{autobreak}
\beta^{(1)}(\mu_H^{2}) =

-  \frac{3}{2} g_1^{2} \mu_H^{2}

-  \frac{9}{2} g_2^{2} \mu_H^{2}

+ 12 \lambda_H \mu_H^{2}

+ \lambda_{HS} \mu_S^{2}

+ 6 \mu_H^{2} y_t^{2}
\end{autobreak}
\end{align*}
\begin{align*}
\begin{autobreak}
\beta^{(2)}(\mu_H^{2}) =

+ \frac{557}{48} g_1^{4} \mu_H^{2}

+ \frac{15}{8} g_1^{2} g_2^{2} \mu_H^{2}

-  \frac{145}{16} g_2^{4} \mu_H^{2}

+ 24 g_1^{2} \lambda_H \mu_H^{2}

+ 72 g_2^{2} \lambda_H \mu_H^{2}

- 60 \lambda_H^{2} \mu_H^{2}

-  \frac{1}{2} \lambda_{HS}^{2} \mu_H^{2}

- 2 \lambda_{HS}^{2} \mu_S^{2}

+ \frac{85}{12} g_1^{2} \mu_H^{2} y_t^{2}

+ \frac{45}{4} g_2^{2} \mu_H^{2} y_t^{2}

+ 40 g_3^{2} \mu_H^{2} y_t^{2}

- 72 \lambda_H \mu_H^{2} y_t^{2}

-  \frac{27}{2} \mu_H^{2} y_t^{4}
\end{autobreak}
\end{align*}
\begin{align*}
\begin{autobreak}
\beta^{(1)}(\mu_S^{2}) =

+ 4 \lambda_{HS} \mu_H^{2}

+ 6 \lambda_S \mu_S^{2}
\end{autobreak}
\end{align*}
\begin{align*}
\begin{autobreak}
\beta^{(2)}(\mu_S^{2}) =

+ 8 g_1^{2} \lambda_{HS} \mu_H^{2}

+ 24 g_2^{2} \lambda_{HS} \mu_H^{2}

- 8 \lambda_{HS}^{2} \mu_H^{2}

- 30 \lambda_S^{2} \mu_S^{2}

- 2 \lambda_{HS}^{2} \mu_S^{2}

- 24 \lambda_{HS} \mu_H^{2} y_t^{2}
\end{autobreak}
\end{align*}

	\bibliographystyle{unsrt}
	\bibliography{references}

\end{document}